\documentclass[10pt,twocolumn,twoside]{IEEEtran}
\usepackage[utf8]{inputenc}
\usepackage{anyfontsize}
\usepackage{array,multirow,graphicx}

\usepackage{cite}

\usepackage{amsmath,amssymb,amsfonts}

\usepackage{subfigure}
\usepackage{scalefnt}

\usepackage{color}
\usepackage[T1]{fontenc}

\begin{document}
\title{Objective Human Affective Vocal Expression Detection and Automatic Classification with 
Stochastic Models and Learning Systems}
\author{V.~Vieira,~\IEEEmembership{Student Member,~IEEE,}
        R.~Coelho,~\IEEEmembership{Senior Member,~IEEE,}
        and~F. M. de~Assis \vspace{-3.5ex}
\thanks{This work was supported in part by the National Council for Scientific and Technological Development (CNPq) under research grants 140816/2014-3 and 307866/2015-7, and Fundação de Amparo à
Pesquisa do Estado do Rio de Janeiro (FAPERJ) under research grant 203075/2016.}        
\thanks{V. Vieira is with the Post-Graduate Program in Electrical Engineering, Federal University of Campina Grande (UFCG), Campina Grande 58429-900, Brazil (e-mail: vinicius.vieira@ee.ufcg.edu.br).}
\thanks{R. Coelho is with the Laboratory of Acoustic Signal Processing (lasp.ime.eb.br), Military Institute of Engineering (IME), Rio de Janeiro 22290-270, Brazil (e-mail: coelho@ime.eb.br).}
\thanks{F. M. de Assis is with the Electrical Engineering Department, Federal University of Campina Grande (UFCG), Campina Grande 58429-900, Brazil (e-mail: fmarcos@dee.ufcg.edu.br).}
}

\maketitle

\begin{abstract}
This paper presents a widespread analysis of affective vocal expression classification systems.
In this study, state-of-the-art acoustic features are compared to two novel affective vocal prints
for the detection of emotional states: 
the Hilbert-Huang-Hurst Coefficients (HHHC) and the vector of index of non-stationarity~(INS). 
HHHC is here proposed as a nonlinear vocal source feature vector that represents the affective states according to their 
effects on the speech production mechanism. Emotional states are highlighted by the empirical mode decomposition (EMD) based method, which exploits the non-stationarity of the affective acoustic variations. Hurst coefficients~(closely related to the excitation source) are then estimated from the decomposition process to compose the feature vector. Additionally, the INS vector 
is introduced as dynamic information to the HHHC feature. 
The proposed features are evaluated in speech emotion classification experiments with three databases in German and English languages. 
Three state-of-the-art acoustic features are adopted as baseline.
The $\alpha$-integrated Gaussian model ($\alpha$-GMM) is also introduced for the emotion representation and classification.
Its performance is compared to competing stochastic and machine learning classifiers. 
Results demonstrate that HHHC leads to significant classification improvement when compared to the 
baseline acoustic features. 
Moreover, results also show that $\alpha$-GMM outperforms the competing classification methods.
Finally, HHHC and INS are also evaluated as complementary features for the GeMAPS and eGeMAPS feature sets.
\end{abstract}

\vspace{-.2cm}
\section{Introduction}
\IEEEPARstart{A}{ffective} states play an important role in the cognition, perception and communication of the human-being daily life.
For instance, an unexpected event can motivate a happiness state. 
On the other hand, stressful situations may cause health problems.
Automatic emotion recognition is especially important to improve communication 
between human and machine \cite{cowie2001emotion,ekman1999}.
In the literature, emotions are generally classified using physical or physiological signals such as 
speech \cite{el2011survey}, facial expression \cite{bar15}, and
electrocardiogram (ECG) \cite{agrafioti2012}.
Particularly, speech emotion recognition has received much research attention in the past few years \cite{tawari2010speech,schuller2009acoustic,Coelho_pH,zhang2018speech}. 
In this scenario, many \mbox{promising} applications can be considered, such as security access, \mbox{automatic} translation, call-centers, mobile \mbox{communication} and human-robot interaction~\cite{tahon2016towards}. 

\begin{figure}[t!]
\centering
\includegraphics[width=0.48\textwidth]{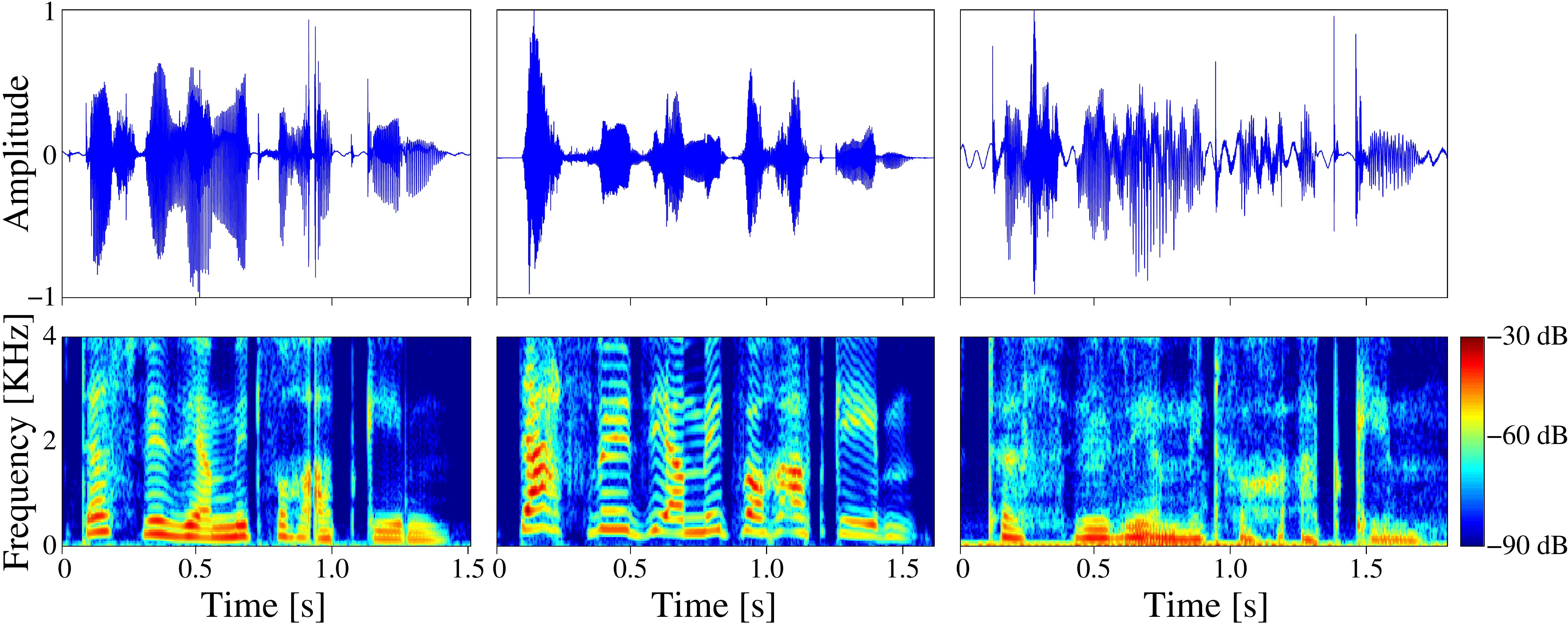}\\
\scriptsize(a)\hspace{0.13\textwidth}(b)\hspace{0.14\textwidth}(c)\hspace{0.05\textwidth}
\vspace{-0.3cm}\caption{Amplitudes and spectrograms of speech signals produced considering different emotional states: 
(a) Neutral, (b) Anger, and (c) Sadness.}
\label{fig:emotions}
\vspace{-0.4cm}
\end{figure}

The speech production under emotions is affected by changes in muscle tension and in the breathing rate.
These changes lead to different speech signals depending on the emotion.
Figure \ref{fig:emotions} depicts amplitudes and corresponding spectrograms of speech signals produced with three affective expressions: Neutral, Anger, and Sadness.
These signals were collected from the Berlin Database of Emotional Speech (EMO-DB) \cite{berlinDatabase} and
were spoken by the same female person and contain the same message.
It can be noted that amplitudes and spectrograms are functions of the affective state.

In the context of social interactions, there is a large number of emotional states~\cite{scherer2003vocal}. According to Ekman~\cite{ekman1999}, there are certain emotions that can be naturally recognized by humans. Although this universality of the affective states discrimination, their decoding in the computational field is difficult. An \textit{affective vocal print} is fundamental to a powerful recognition system. Thus, a key challenge is to define a feature that characterizes different emotions~\cite{el2011survey,tahon2016towards}.  In the literature, there is not yet a consensus about an effective acoustic feature for this task. In this sense, the choice of an attribute that shows meaningful information related to the physiological behavior of multiple affective states is a crucial search. 

In \cite{zhou2001nonlinear},
Teager-Energy-Operator~(TEO)~\cite{teager1980some} based 
features were proposed for the classification of stress conditions. 
The idea was to capture nonlinear airflow structures of the acoustic signal induced by the speaker emotional state.
Based on the fact that the excitation source signal reflects the speaker physiological behavior, 
vocal source features may also be applied for this purpose. Such features are less dependent on the linguistic content of speech~\cite{wang2011robust}, in comparison to spectral ones. In~\cite{Coelho_pH}, the pH vocal source feature~\cite{sant2006text} was evaluated for emotion and stress classification. The authors showed that TEO features may be not suitable for emotion classification. Both pH and TEO features do not take into account the nonlinear effect of the speech production such as the non-stationarity of the affective acoustic variation and its dynamic behavior. These aspects are important to be exploited by an acoustic affective attribute.

One of the most common features applied as baseline in the literature and challenges is the mel-frequency \mbox{cepstral} \mbox{coefficients~(MFCC)}. This feature has been widely used for affective recognition due its success in other tasks, such as speech and speaker  recognition \cite{wang2011robust,wu2011automatic}. Nonetheless, other proposed features have shown superior performance than MFCC~\cite{Coelho_pH,zhou2001nonlinear,wu2011automatic,wang2015speech}. For instance, the Hurst vector (pH) \cite{sant2006text} achieves accuracy 6.8~percentage points~(p.p.) higher than MFCC in emotions classification~\cite{Coelho_pH}.
Some approaches have focused in recognition rates improvement, where several features are combined to form collections of low-level descriptors~(LLDs)~\cite{tahon2016towards,eyben2016geneva}. This means that there is not yet a pure and established attribute for emotion classification. Furthermore, such studies are applied in the context of arousal and valence classification. Additionally, the scope of this present study is the representation of each affective state individually, which can improve the performance of classification tasks.

This work introduces a new nonlinear acoustic feature based on non-stationary effects of emotions. 
The empirical mode decomposition~(EMD)~\cite{huang_98} is applied to emphasize acoustic variations
present in the speech signal. 
Hurst coefficients~\cite{hurst1951long} are then estimated to characterize highlighted vocal source components. 
Finally, the Hilbert-Huang-Hurst Coefficients (HHHC) compose the affective vector on a frame-basis feature extraction. 
The combination of EMD with Hurst exponent is able to capture the non-stationary acoustic variations that occur during 
the speech production depending on the affective states.
This aspect is still not well explored in the literature. 

The index of non-stationarity~(INS)~\cite{flan10} is here proposed as additional information to the HHHC feature vector. It dynamically describes the non-stationary behavior of affective speech samples. The $\alpha$-GMM~\cite{wu2009alpha} 
is also introduced to classify emotional states. It is compared to classic Gaussian Mixture Models (GMM)~\cite{reynolds1995robust} and Hidden Markov Models (HMM)~\cite{rabiner1986introduction} stochastic methods, and also machine learning approaches: Support Vector Machines (SVM)~\cite{cortes1995machine}, Deep Neural Networks (DNN) \cite{deep-learning-2014}, 
Convolutional Neural Networks (CNN) \cite{lecun1998}, and Convolutional Recurrent Neural Networks (CRNN) \cite{cakir2017}). Experiments show the effectiveness of the new vocal source feature in different languages and scenarios. Several results demonstrate that HHHC is a 6-dimensional vector with robustness as a pure attribute for emotion. Additionally, HHHC contributes as complementary to GeMAPS and eGeMAPS\cite{eyben2016geneva} features sets to improve the classification rates.

This paper is organized as follows. Section \ref{sec:hhhc} introduces the HHHC feature and presents the feature extraction procedure. 
The INS is also described in this section. The $\alpha$-GMM and competing classifiers are presented in Section \ref{sec:classifier}. 
Evaluation experiments are described in Section \ref{sec:exp} and results are exhibited in Section 
\ref{sec:result}. 
Finally, Section \ref{sec:conclusion} concludes this work.

\begin{figure*}[t!]
\centering
\includegraphics[width=.3\textwidth]{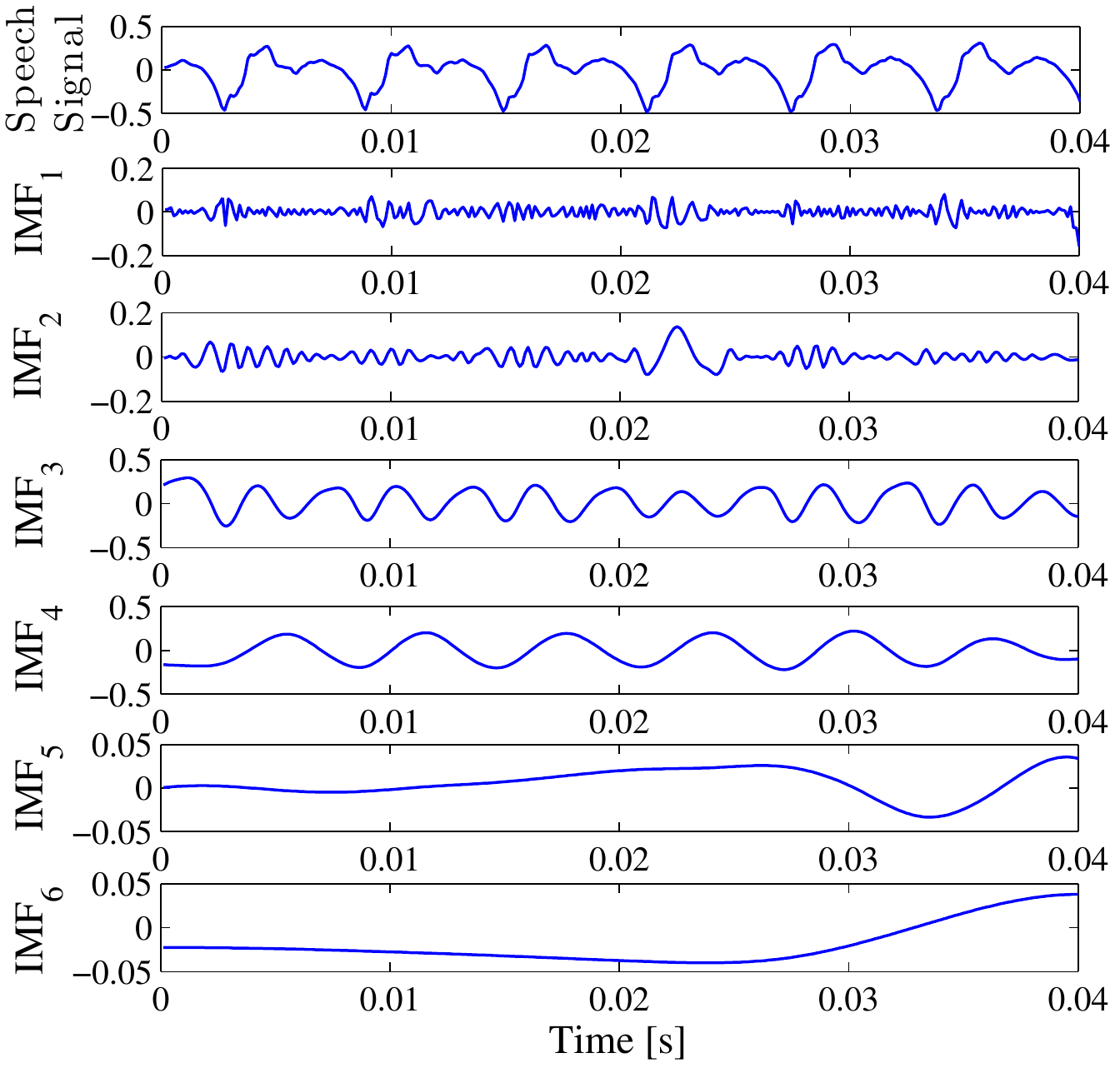}\hspace{.04\textwidth}
\includegraphics[width=.3\textwidth]{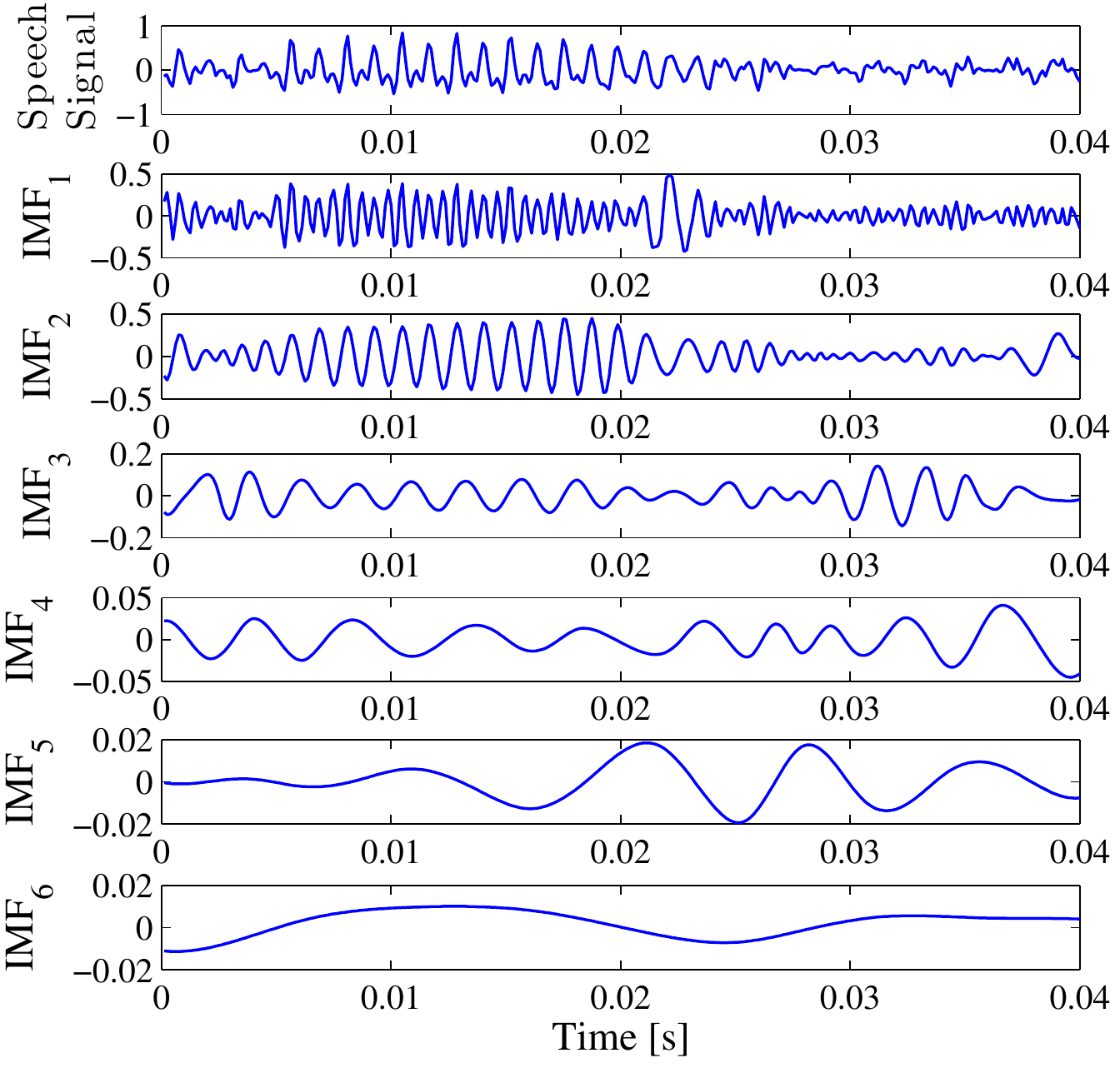}\hspace{.04\textwidth}
\includegraphics[width=.3\textwidth]{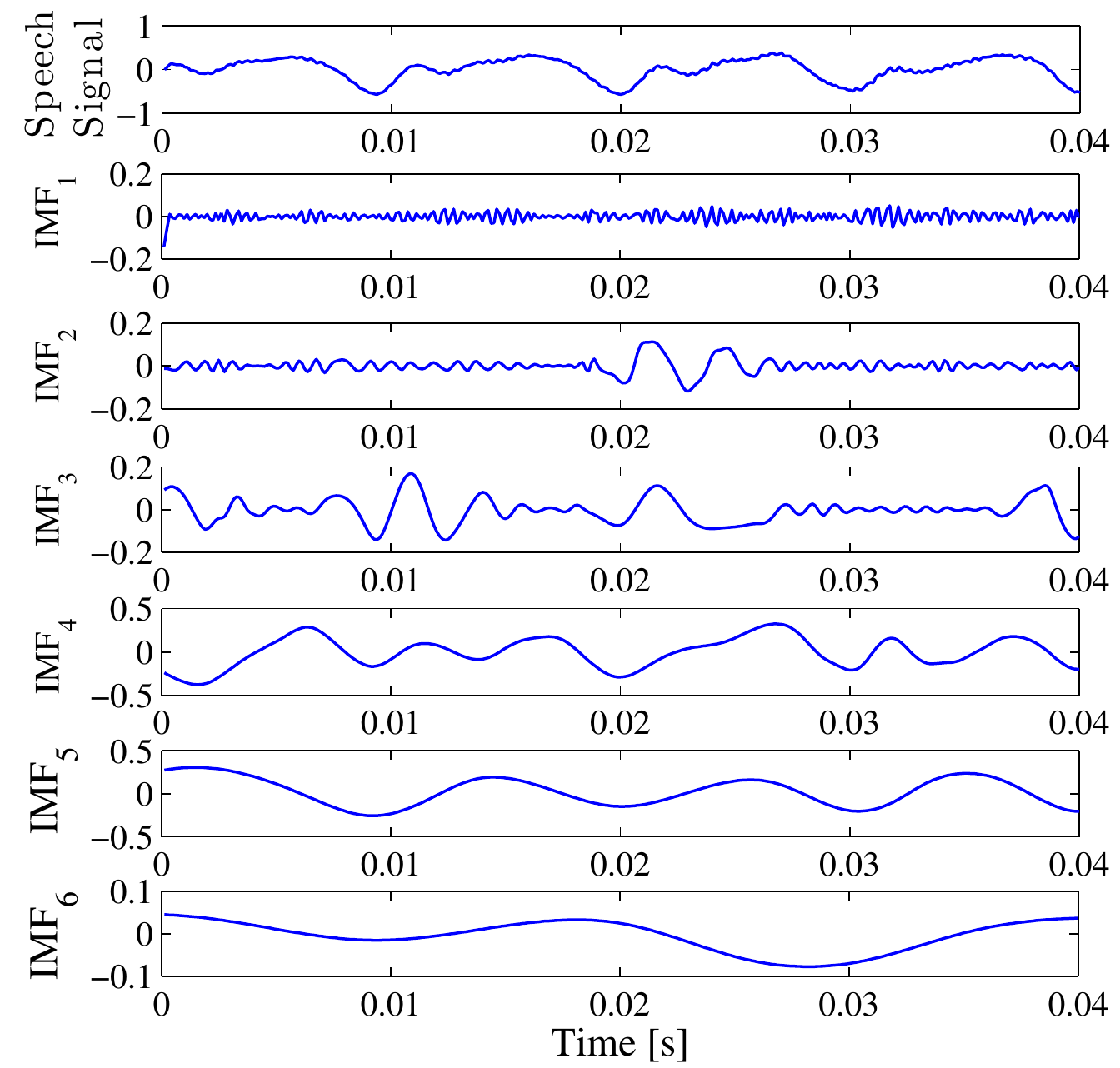} \\
\scriptsize\hspace{0.02\textwidth}(a)\hspace{0.33\textwidth}(b)\hspace{0.33\textwidth}(c)\\
\vspace{-0.2cm}
\caption{First six IMFs obtained with EEMD from voiced speech segments: 
(a) Neutral, (b) Anger, and (c) Sadness.}
\label{fig:emdeemd}
\vspace{-0.5cm}
\end{figure*}

\vspace{-.2cm}
\section{A New Nonlinear Acoustic Feature}
\label{sec:hhhc}

The general idea of the Hilbert-Huang-Hurst Coefficients (HHHC) vector is to characterize the vocal source when affected by an emotional state. The affective content of the speech is highlighted by an adaptive method based on Hilbert-Huang transform~(EMD). Instead of the original EMD, the ensemble EMD (EEMD)~\cite{huang09} is applied to analyze an improvement in the affective states detection. After the decomposition, Hurst coefficients, which are related to the excitation source, capture the nonlinear information from the emphasized acoustic variations. 
In~\cite{venturini2014speech}, 
it was shown that acoustic sources have different degrees of non-stationarity. In this work, a vector of INS values is proposed to analyze and detect speech emotional states.

\vspace{-.2cm}
\subsection{HHHC Feature}

The HHHC vocal source feature is obtained by using the EMD-based approach and the estimation of Hurst coefficients from the decomposition process.

\subsubsection{EMD/EEMD}
EMD was introduced in~\cite{huang_98} as a nonlinear time-domain adaptive method for decomposing non-stationary signals into a series of oscillatory modes. The general idea is to locally analyze a signal $x(t)$ between two consecutive extrema~(minima or maxima). Then, two parts are defined: a local fast component, also called detail, $d(t)$, and the local trend or residual $a(t)$, 
such that $x(t)=d(t)+a(t)$. The detail function $d(t)$ corresponds to the first intrinsic mode function (IMF) and consists of the highest frequency component of $x(t)$. The subsequent IMFs are iteratively obtained from the residual of the previous IMF. 
The decomposition can be summarized in the following steps:

\begin{enumerate}
\item Identify all local extrema (minima and maxima) of $x(t)$;

\item Interpolate the local maxima and minima via cubic splines to obtain the upper ($e_{up}(t)$) and lower ($e_{lo}(t)$) envelopes, respectively;

\item Define the local trend as 
$a(t)=~(e_{up}(t)~+~e_{lo}(t))~/~2$;

\item Calculate the detail component as $d(t)=x(t)-a(t)$.
\end{enumerate}

Every IMF have zero mean, and the numbers of maxima and zero-crossings must be equal or differ by at most one. 
If the detail component $d(t)$ does not follow these properties, steps 1-4 are repeated with $d(t)$ in place of $x(t)$ until the new detail can be considered as an IMF. For the next IMF, the same procedure is applied on the residual $a(t)=x(t)-d(t)$.

Since an input signal $x(t)$ can be decomposed in a finite number of IMFs, the integrability property of the EMD can be expressed as
$x(t) = \sum_{m=1}^{M} \mbox{IMF}_m(t) + r(t)$,
where $r(t)$ is the last residual sequence.


As an alternative for EMD, the EEMD method was proposed to avoid the \textit{mode mixing}
phenomena~\cite{huang09}, which refer to IMF fluctuations that do not appear in the proper scale. Thus, the EEMD approach is expected to emphasize affective acoustic variations.
Given the target signal $x(t)$, the EEMD method firstly generates an ensemble of $I$ trials, $x^{i}(t)$, $i=1,...,I$, each consisting of $x(t)$ plus a white noise of finite amplitude, $w^{i}(t)$, i.e.,
$x^{i}(t)=x(t)+w^{i}(t)$.
Each trial $x^{i}(t)$ is decomposed with EMD leading to $M$ modes, $\mathrm{IMF}^{i}_{m}(t)$, $m=1,...,M$. Then, the $m$-th mode of $x(t)$ is obtained as the average of the $I$ corresponding IMFs.


Figure \ref{fig:emdeemd} shows the EEMD applied to three speech segments of 400~ms collected from 
EMO-DB \cite{berlinDatabase}.
Segments refer to Neutral speech (Figure \ref{fig:emdeemd}a) and two basic emotions: Anger (Figure \ref{fig:emdeemd}b) and Sadness (Figure \ref{fig:emdeemd}c). The EEMD applies a high-frequency versus low-frequency separation between IMFs. Note that the affective signals have different non-stationary dynamic behaviors. For instance, IMFs 1 and 2 for Anger present amplitude values higher than for the other signals. On the other hand, the highest amplitude values are observed in the late three oscillations~(IMFs 4, 5 and 6) of the Sadness state. This indicates that EEMD highlights the affective content of speech. For high-arousal emotions (e.g., Anger), non-stationary acoustic variations are more concentrated in the high-frequency IMFs, while the low-frequency ones capture the prevailing content from the low-arousal emotions (e.g., Sadness).

\subsubsection{Hurst Coefficients}
The Hurst exponent ($0<H<1$), or Hurst coefficient, expresses the time-dependence or scaling degree of a stochastic process~\cite{hurst1951long}. Let a speech signal be represented by a stochastic process $x(t)$, with the normalized autocorrelation coefficient function $\rho(k)$, the $H$ exponent is defined by the asymptotic behavior of $\rho(k)$ as $k\rightarrow \infty$, i.e.,
$\rho(k) \sim H(2H-1)k^{2(H-2)}$.

In this study, the $H$ values are estimated from IMFs on a frame-by-frame basis using the wavelet-based estimator~\cite{veitch1999wavelet}, which can be described in three main steps as follows:
\begin{enumerate}
\item Wavelet decomposition: the discrete wavelet transform~(DWT) is applied to successively decompose the input sequence of samples into approximation~($a_{w}(j,n)$) and detail~($d_{w}(j,n)$) 
coefficients
, where $j$ is the decomposition scale~($j=1,2,...,J$) and $n$ is the coefficient index of each scale.

\item Variance estimation: for each scale $j$, the variance $\sigma^{2}=(1/N_{j})\sum_{n}d_{w}(j,n)^{2}$ is evaluated from detail coefficients, where $N_{j}$ is the number of available coefficients for each scale $j$. In~\cite{veitch1999wavelet}, it is shown that $E[\sigma_{j}^{2}]=C_{H}j^{2H-1}$, where $C_{H}$ is a constant.

\item Hurst computation: a weighted linear regression is used to obtain the slope $\theta$ of the plot of $y_{i}=\mathrm{log}_{2}(\sigma_{j}^{2})$ versus $j$. The Hurst exponent is estimated as $H=(1+ \theta)/2$.
\end{enumerate}

\begin{figure}[!t]
\centering
\includegraphics[width=0.4\textwidth]{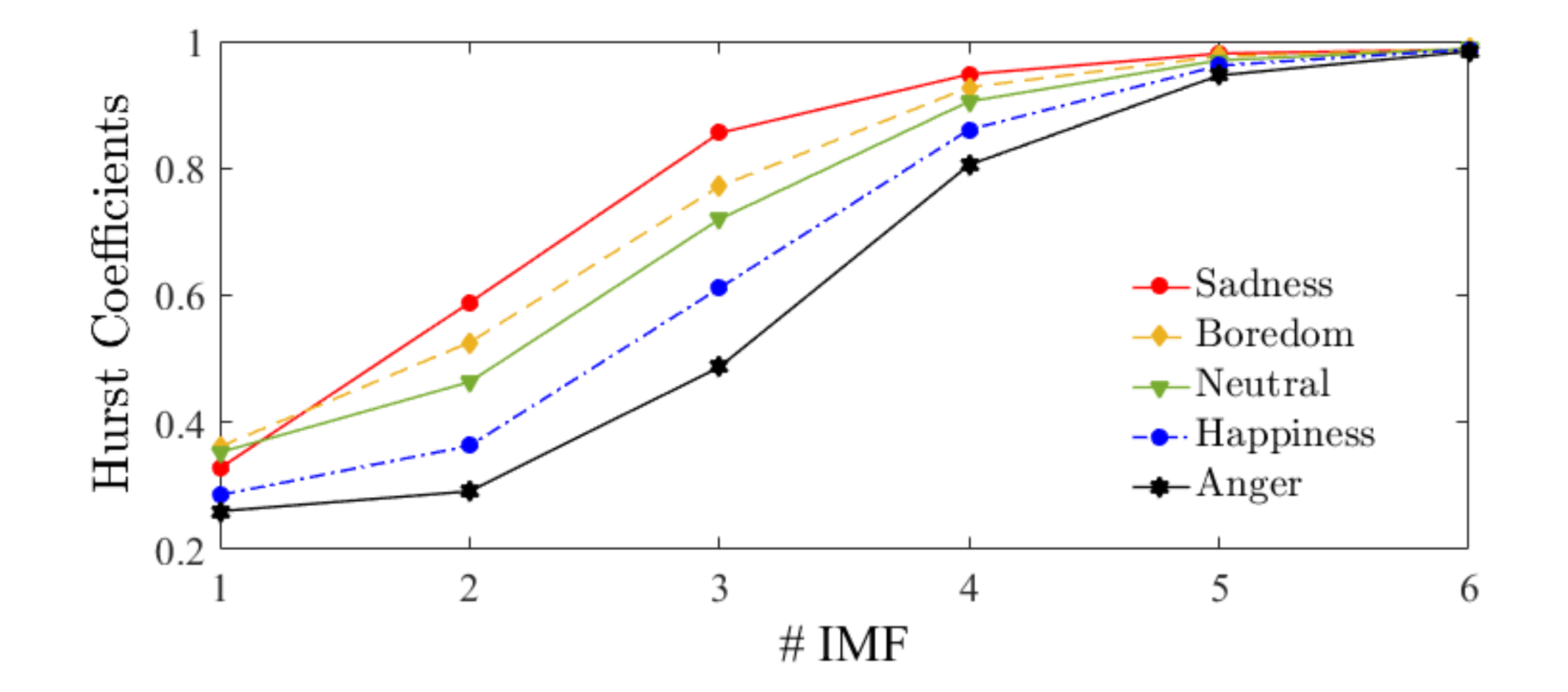} 
\vspace{-0.3cm}
\caption{\label{fig:medias}Hurst mean values of six IMFs obtained from speech samples under five non-stationary emotional variations.}
\vspace{-0.3cm}
\end{figure}

In~\cite{Coelho_pH}, it was shown that $H$ is related to the excitation source of emotional states. A high-arousal emotional signal has $H$ values close to zero, while a low-arousal one has $H$ values close to the unity. The authors extracted Hurst coefficients directly from the speech signal in a frame-basis for the pH feature~\cite{Coelho_pH}. In contrast, this present work deals with the estimation of Hurst values from IMFs of speech signals.

The HHHC vector for speech samples is illustrated in Figure \ref{fig:medias}. Signals are collected from the EMO-DB corresponding to five different emotional variations: Sadness, Boredom, Neutral, Happiness and Anger. A time duration of 40 s is considered for each emotional state. Six IMFs are obtained by the EEMD method, applied to speech segments of 80~ms and 50\% overlapping. The Hurst exponent is computed and averaged from non-overlapping frames of 20 ms within each IMF, using Daubechies filters~\cite{daubechies1992ten} with 12 coefficients and 3-12 scales in the wavelet-based Hurst estimator. It can be seen that the vocal source featured by Hurst coefficients are highlighted by the EEMD. Note that low-arousal emotions present the highest $H$ values for the majority of the IMFs. For all the analyzed IMFs, high-arousal emotions have the lowest $H$ averages.

\begin{figure}[!t]
\centering
\includegraphics[width=0.49\textwidth]{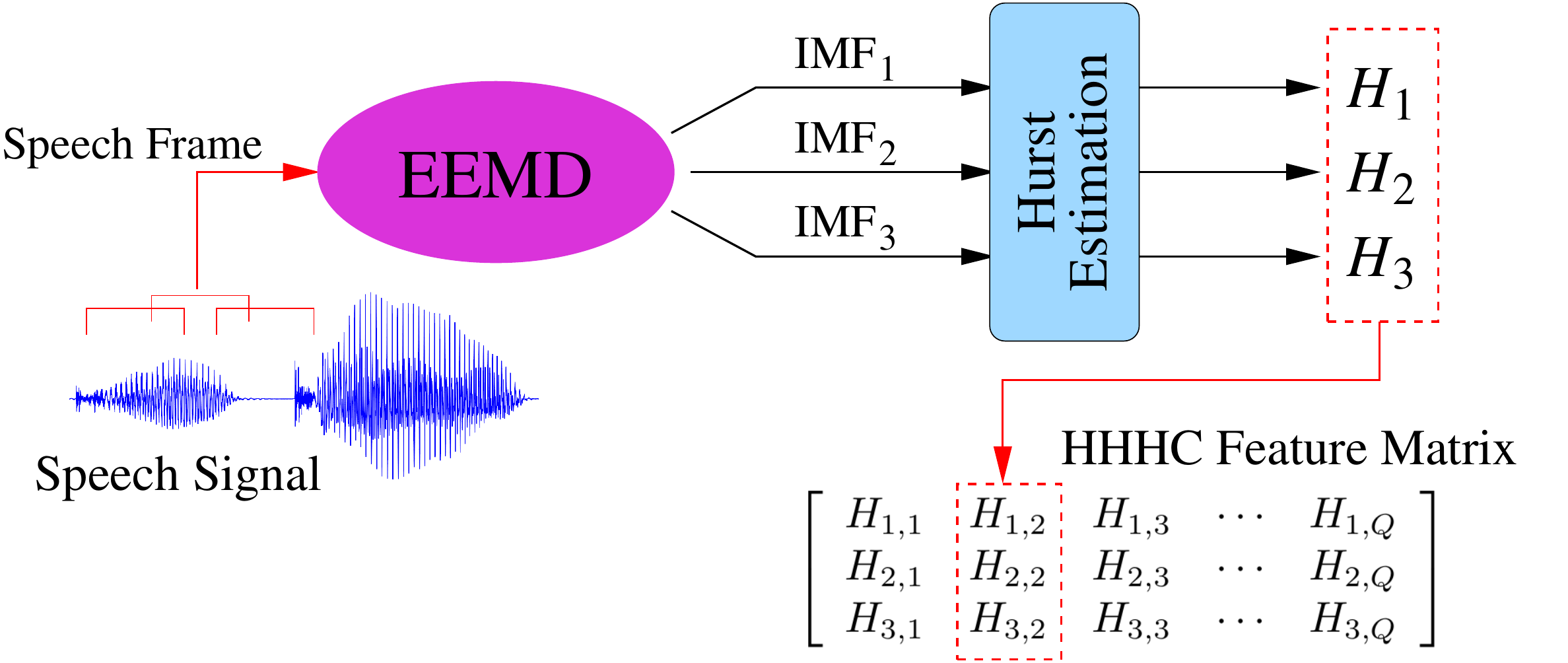}
\vspace{-0.3cm}
\caption{\label{fig:HHHC}An example of a HHHC vector extraction with 3 coefficients.}
\vspace{-0.3cm}
\end{figure}

\subsubsection{HHHC Feature Extraction}
The HHHC extraction of affective speech signals is performed in two main steps: signal decomposition using EMD or EEMD; and multi-channel estimation of the Hurst exponent. An example of the HHHC vector estimation with 3 values of $H$ is presented in Figure \ref{fig:HHHC}. The decomposition is applied to each segment of the input signal. The Hurst coefficients are obtained in a frame-by-frame basis from each IMF. 
Then, the feature matrix for HHHC is formed as an acoustic feature.

\vspace{-.2cm}
\subsection{INS Vector}

The INS is a time-frequency approach to objectively \mbox{examine} the non-stationarity of a signal~\cite{flan10}. The stationarity test is conducted by comparing spectral components of the signal to a set of stationary references, called \textit{surrogates}. For this purpose, spectrograms of the signal and surrogates are obtained by means of the short time Fourier transform (STFT).
Then, the Kullback-Leibler (KL) divergence is used to measure the distance between the spectrum of the analyzed signal and its global spectrum averaged over time. Given KL$^{(x)}$ for the analyzed signal $x(t)$ and KL$^{(s_{j})}$ for the $j$ surrogates obtained from $x(t)$. Since there are $N$ short spectrograms, a variance measure, $\Theta$, is obtained from the
KL values:
\begin{equation}
\left\{
\begin{array}{ll}
\Theta_0(j) = \mathrm{var}\left(\mathrm{KL}_{n}^{(s_{j})} \right)_{n=1,...,N.}, \hspace{4mm} j=1,...,J. \\
 \Theta_1 = \mathrm{var}\left(\mathrm{KL}_{n}^{(x)} \right)_{n=1,...,N.} 
\end{array}\right.
\label{eq:ins22}
\end{equation}
Finally, the INS is given by
$\mathrm{INS} := \sqrt {{\Theta_1}/{\langle \Theta_{0}(j) \rangle_{j}}}$,
where $\langle \cdot \rangle$ is the mean value of $\Theta_{0}(j)$. In~\cite{flan10}, the authors considered that the distribution of the KL values can be approximated by a Gamma distribution. Therefore, for each window length $T_{h}$, a threshold $\gamma$ can be defined for the stationarity test considering a confidence degree of 95\%. Thus,
\begin{equation}
\mbox{INS} \left\{
\begin{array}{ll}
\leq \gamma &\mbox{, signal is stationary;}\\
> \gamma & \mbox{, signal is non-stationary.}
\end{array}\right.
\label{eq:ins}
\end{equation}

\begin{figure}[!t]
\centering
\includegraphics[width=0.49\textwidth]{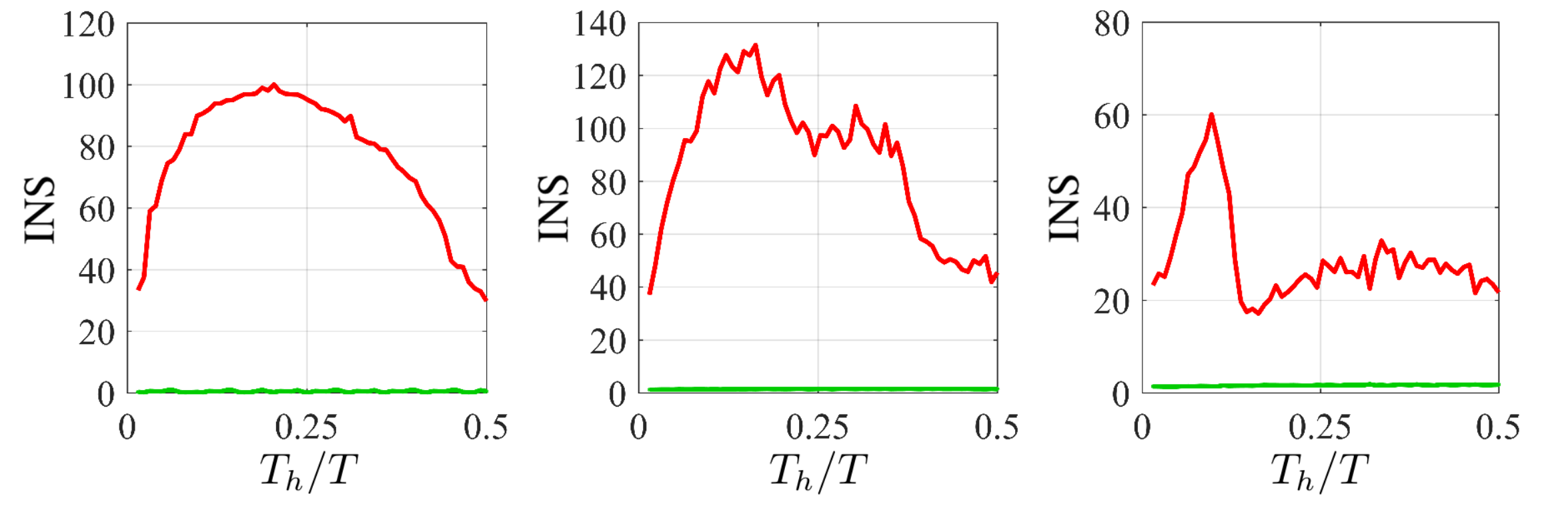}\\ 
\vspace{-0.1cm}
\scriptsize\hspace{0.3cm} (a) \hspace{2.4cm} (b) \hspace{2.4cm} (c) \\
\vspace{-0.2cm}
\caption{\label{fig:ins}INS computed from voiced segments considering emotional states: (a) Neutral, (b) Anger and (c) Sadness.}
\vspace{-0.3cm}
\end{figure}

Figure \ref{fig:ins} depicts examples of the INS obtained from voiced segments of the Neutral state and two emotional variations: Anger and Sadness. The time scale $T_{h}/T$ is the ratio between the length adopted in the short-time spectral analysis~($T_{h}$) and the total length ($T=$ 800 ms) of the signal. Note that INS for both emotional states (red line) is higher than the threshold adopted in the test of non-stationarity (green line). However, the INS values vary from one emotional state to another. While the Neutral state has INS values in the range~[50,100] for the majority of the observed time-scales, the INS for Sadness reaches a maximum value of 60. On the other hand, Anger presents INS greater than 100 for several time-scales.

\vspace{-.2cm}
\section{Classification Task}
\label{sec:classifier}
The $\alpha$-integrated Gaussian Mixture Model is here proposed for acoustic emotion classification. The $\alpha$-GMM was firstly proposed for speaker identification \cite{wu2009alpha}. By introducing a factor of $\alpha$, the modelling capacity of the GMM is extended, which is more suitable in acoustic variations conditions. 
The $\alpha$-integration generalizes the linear combination adopted in the conventional GMM ($\alpha=-1$). For $\alpha < -1$, the $\alpha$-GMM classifier emphasizes larger probability values and de-emphasizes smaller ones. Since affective states are assumed as acoustic variations added to speech in its production, it is understood that $\alpha$-GMM increases the recognition performance.  Similar to what was shown in~\cite{wu2009alpha}, it was demonstrated in~\cite{venturini2014speech} that \mbox{$\alpha$-GMM} outperforms the conventional GMM. Hence, the HHHC is  evaluated considering the $\alpha$-GMM and the classical GMM ($\alpha=-1$). Five other classifiers are used for comparative evaluation purposes.

\vspace{-.2cm}
\subsection{$\alpha$-integrated Gaussian Mixture Model ($\alpha$-GMM)}

Given an affective state model $\lambda_{L}$, composed of $M$ Gaussian densities $b_{i}(\mathrm{\textbf{x}})$, $i=1,...,M$, the $\alpha$-integration of densities is defined as~\cite{wu2009alpha},
\begin{equation}
p \left( \mathrm{\textbf{x}} | \lambda_{L} \right) = C \left[ \sum_{i=1}^{M} \pi_{i}   b_{i}(\mathrm{\textbf{x}})^{\frac{1-\alpha}{2}} \right]^{\frac{2}{1-\alpha}} ,
\label{eq:gmm4}
\end{equation}
where $\pi_{i}$ are non-negative mixture weights constrained to $\sum_{i=1}^{M} \pi_{i}=1$, and $C$ is a normalization constant. Note that $\alpha=-1$ corresponds to the conventional GMM.

Models $\lambda_{L}$ are completely parametrized by mean vectors, covariance matrices, and weights of Gaussian densities. These parameters are estimated using an adapted expectation-maximization~(EM) algorithm as to maximize the likelihood function
$p \left( \mathrm{\textbf{X}} | \lambda_{L} \right) = \prod_{t=1}^{Q} p \left( \mathrm{\textbf{x}}_{t} | \lambda_{L} \right)$,
where $\mathrm{\textbf{X}} = \left[ \textbf{x}_1 \textbf{x}_2 \ldots \textbf{x}_Q \right]$ is the feature matrix extracted from the training speech segment $\Phi_{L}$ of the affective state $L$.

\vspace{-.2cm}
\subsection{Hidden Markov Models (HMM)}

The HMM consists of finite internal states that generate a set of external events (observations). These states are hidden for the observer, and capture the temporal structure of an affective speech signal. \mbox{Mathematically}, the HMM can be characterized by three fundamental problems:

\begin{enumerate}
\item Likelihood: Given an HMM $\lambda_{L} = (A,B)$ with $K$ states, and an observation sequence $\mathrm{\textbf{x}}$, determine the likelihood $p(\mathrm{\textbf{x}}|\lambda_{L})$, where $A$ is a matrix of transitions probabilities $a_{jk}$, $j,k=1,2,...,K$, from state $j$ to state $k$, and  $B$ is the set of densities $b_{j}$;

\item Decoding: Given an observation sequence $\mathrm{\textbf{x}}$ and an HMM $\lambda_{L}$, discover the sequence of hidden states;

\item Learning: Given an observation sequence $\mathrm{\textbf{x}}$ and the set of states in the HMM, learn the parameters $A$ and $B$.
\end{enumerate}

The standard algorithm for HMM training is the forward-backward, or Baum-Welch algorithm~\cite{baum1972inequality}. It obtains $A$ and $B$ matrices which maximizes the likelihood $p(\mathrm{\textbf{x}}|\lambda_{L})$. The Viterbi algorithm is commonly used for decoding~\cite{viterbi1967error}.

\vspace{-.2cm}
\subsection{Support Vector Machines (SVM)}

SVM~\cite{cortes1995machine} is a classical supervised machine learning model widely applied for data classification. The general idea is to find  the optimal separating 
hyperplane which maximizes the margin on the training data. For this purpose, it transforms input vectors into a high-dimensional feature space using a nonlinear transformation~(with a kernel function). 
Given a training set $\lbrace u_{\xi}\rbrace_{\xi=1}^{N} = \lbrace \left( \mathrm{\textbf{x}}_{\xi}, L_{\xi} \right) \rbrace_{\xi=1}^{N}$, where $L_{\xi} \in \lbrace -1, +1 \rbrace$ represents the affective state $L$ of the utterance $\xi$. Thus, the classifier is a hyperplane defined as
$g(\mathrm{\textbf{x}})= \textbf{\textit{w}}^{\mathrm{T}} \mathrm{\textbf{x}} +b$,
where $\textbf{\textit{w}}$ is the gradient vector which is perpendicular to the hyperplane, and $b$ is the offset of the hyperplane from the origin. The side of the hyperplane which belongs the utterance can be indicated by $L_{\xi}g(\mathrm{\textbf{x}}_{\xi})$. For $L_{\xi}=+1$, $L_{\xi}g(\mathrm{\textbf{x}}_{\xi})$ must be greater than $1$, while $L_{\xi}g(\mathrm{\textbf{x}}_{\xi})$ is required to be smaller than $-1$ for $L_{\xi}=-1$. Then, the hyperplane is chosen by the solution of the optimization problem of minimizing $\frac{1}{2}\textbf{\textit{w}}^{\mathrm{T}}\textbf{\textit{w}}$ subject to 
$L_{\xi} \left( \textbf{\textit{w}}^{\mathrm{T}} \mathrm{\textbf{x}} +b \right) \geq 1, \xi=1,2,...,N.$

In this work, the input data for the SVM classifier is obtained from mean vectors of feature matrices. This statistic was more prominent than others, such as median and maximum value, as observed in~\cite{milton2013svm}. Radial Basis Function~(RBF) is used as the SVM kernel.

\vspace{-.2cm}
\subsection{Deep Neural Networks (DNN)}
DNN is one of the most prominent methods for machine learning tasks such as 
speech recognition~\cite{hinton2012deep}, separation~\cite{wang2014training}, and emotion classification \cite{zhang2018speech}. 
The deep learning concept can be applied for architectures such as feedfoward multilayer perceptrons~(MLPs), convolutional neural networks~(CNNs) and recurrent neural networks~(RNNs)~\cite{wang2017supervised}.
In this work, it is considered MLP that has feedforward connections from the input layer to the output layer, with sigmoid activation function $y_{j}$ for the neuron $j$,
$y_{j}={1}/(1+ e^{-x_{j}})$,
where $x_{j}=b_{j} + \sum_{i} y_{i}w_{ij}$ is a weighted sum of the previous neurons with a bias $b_{j}$~\cite{hinton2012deep}. 

\vspace{-.2cm}
\subsection{Convolutional Neural Networks (CNN)}

Convolutional Neural Networks \cite{lecun1998} have been widely adopted in the acoustic signal processing area, 
particularly for sound classification \cite{piczak2015,salamon2017} and sound event detection \cite{zhang2015}. 
CNNs extend the multilayer perceptrons model by the introduction of a group of convolutional and pooling layers.
The convolutional kernels are proposed to better capture and classify the spectro-temporal patterns of acoustic signals.
Pooling operations are then applied for dimensionality reduction between convolutional layers.

\vspace{-.2cm}
\subsection{Convolutional Recurrent Neural Networks (CRNN)}

CRNNs \cite{cakir2017} consist on the combination of CNNs with Recurrent Neural Networks (RNN). 
The idea is to improve the CNN by learning spectro-temporal information of relatively longer events 
that are not captured by the convolutional layers.
For this purpose, recurrent layers are applied to the output of the convolutional layers 
to integrate the information of earlier time windows.
In the literature, CNNs and RNNs have been successfully combined for music classification \cite{choi2017} 
and sound event detection \cite{cakir2017}.
In this work, a single feedforward layer with sigmoid activation function that follows the recurrent layers 
is considered as the output layer of the network \cite{cakir2017}.

\vspace{-.2cm}
\section{Experimental Setup}
\label{sec:exp}

\begin{figure}[t!]
\centering
\includegraphics[width=0.47\textwidth]{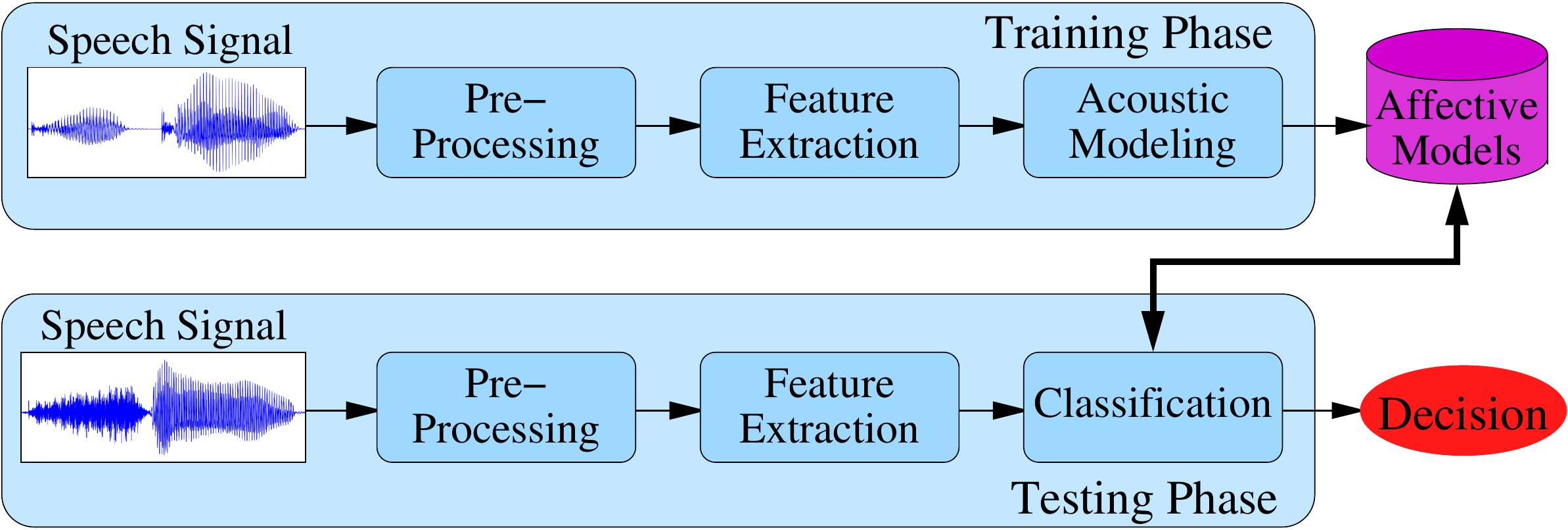} 
\vspace{-0.2cm}
\caption{\label{fig:modelo}Affective vocal expression: classification system diagram.}
\vspace{-0.2cm}
\end{figure}

\begin{table*}[!ht]
\renewcommand{\arraystretch}{1.05}
\centering 
\fontsize{7}{7}\selectfont
\vspace{-.2cm}
\setlength{\tabcolsep}{2.5pt}
\caption{Accuracy Rates (\%) of 5 Emotional States with the HHHC and baseline features for EMO-DB.}
\vspace{-.4cm}
\begin{center}
\begin{tabular}{|c||c|c|c|c|c|c||c|c|c|c|c||c|c|c|c|c||c|c|c|c|c||c|c|c|c|c|}
\cline{3-27} 
\multicolumn{2}{c|}{\rule{0pt}{7pt}} & \multicolumn{5}{c||}{HHHC feature}  & \multicolumn{5}{c||}{HHHC + INS}  
& \multicolumn{5}{c||}{pH feature} & \multicolumn{5}{c||}{MFCC feature} & \multicolumn{5}{c|}{TEO feature}\\ 
\hline
\multirow{8}{*}{\rotatebox[origin=c]{90}{$\alpha$-GMM Classifier }} & \rule{0pt}{7pt}Actual & \multicolumn{5}{c||}{Classified Emotion}  
& \multicolumn{5}{c||}{Classified Emotion}  & \multicolumn{5}{c||}{Classified Emotion} & \multicolumn{5}{c||}{Classified Emotion} & \multicolumn{5}{c|}{Classified Emotion} \\  
 & Emotion & Ang. & Hap. & Neu. & Bor. & Sad.  & Ang. & Hap. & Neu. & Bor. & Sad.  & Ang. & Hap. & Neu. & Bor. & Sad. & Ang. & Hap. & Neu. & Bor. & Sad.  & Ang. & Hap. & Neu. & Bor. & Sad.  \\ \cline{2-27}  
 & \rule{0pt}{7pt} Anger 	& \textbf{86}  &  14 & 0 & 0 & 0   	& \textbf{88}  &  12 & 0 & 0 & 0   	& \textbf{82}  &  18 & 0 & 0 & 0   	& \textbf{80}  &  20 & 0 & 0 & 0   	& \textbf{43}  &  41 & 16 & 0 & 0   \\ 
 & Happiness 				& 35 &	\textbf{65}  & 0 & 0 & 0  	& 32 &	\textbf{68}  & 0 & 0 & 0  	& 41 &	\textbf{55}  & 4 & 0 & 0  	& 18 &	\textbf{80}  & 2 & 0 & 0  	& 31 &	\textbf{55}  & 10 & 4 & 0  	\\ 
 & Neutral 					& 0 & 0 & \textbf{86}  & 14 & 0   	& 0 & 0 & \textbf{87}  & 13 & 0   	& 0 & 6 & \textbf{69}  & 14 & 11   	& 0 & 17 & \textbf{55}  & 19 & 9   	& 8 & 18 & \textbf{47}  & 27 & 0   	\\ 
 & Boredom 					& 0  & 0  & 14 & \textbf{71}  & 15  & 0  & 0  & 10 & \textbf{77}  & 13  & 0  & 4  & 20 & \textbf{43}  & 33  & 0  & 6  & 30 & \textbf{35}  & 29  & 6  & 14  & 24 & \textbf{43}  & 13 \\ 
 & Sadness 					& 0  & 0 & 0 & 12 & \textbf{88}  	& 0  & 0 & 0 & 11 & \textbf{89}  	& 0  & 2 & 8 & 12 & \textbf{78}  	& 0  & 2 & 8 & 22 & \textbf{68}  	& 4  & 0 & 6 & 14 & \textbf{76}  	\\ \cline{3-27} 
&  & \multicolumn{5}{c||}{\rule{0pt}{7pt} Average: \textbf{79.2}}   & \multicolumn{5}{c||}{Average: \textbf{81.8}}  & \multicolumn{5}{c||}{Average: \textbf{65.4}}  & \multicolumn{5}{c||}{Average: \textbf{63.6}}  & \multicolumn{5}{c|}{Average: \textbf{52.8}}\\ \hline
\hline
\multirow{8}{*}{\rotatebox[origin=c]{90}{HMM Classifier}} & \rule{0pt}{7pt}Actual & \multicolumn{5}{c||}{Classified Emotion}  
& \multicolumn{5}{c||}{Classified Emotion}  & \multicolumn{5}{c||}{Classified Emotion} & \multicolumn{5}{c||}{Classified Emotion} & \multicolumn{5}{c|}{Classified Emotion} \\  
 & Emotion & Ang. & Hap. & Neu. & Bor. & Sad.  & Ang. & Hap. & Neu. & Bor. & Sad.  & Ang. & Hap. & Neu. & Bor. & Sad. & Ang. & Hap. & Neu. & Bor. & Sad.  & Ang. & Hap. & Neu. & Bor. & Sad.  \\ \cline{2-27}  
 & \rule{0pt}{7pt} Anger 	& \textbf{76}  &  24 & 0 & 0 & 0    & \textbf{77}  &  23 & 0 & 0 & 0    & \textbf{78}  &  22 & 0 & 0 & 0    & \textbf{74}  &  24 & 2 & 0 & 0    & \textbf{28}  &  52 & 20 & 0 & 0 	\\ 
 & Happiness 				& 33 &	\textbf{67}  & 0 & 0 & 0   	& 30 &	\textbf{70}  & 0 & 0 & 0   	& 32 &	\textbf{64}  & 4 & 0 & 0 	& 25 &	\textbf{70}  & 5 & 0 & 0   	& 31 &	\textbf{59}  & 5 & 5 & 0   	\\ 
 & Neutral 					& 0 & 0 & \textbf{81}  & 19 & 0    	& 0 & 0 & \textbf{84}  & 16 & 0    	& 0 & 6 & \textbf{64}  & 20 & 10    & 0 & 19 & \textbf{48}  & 23 & 10   & 10 & 34 & \textbf{24}  & 32 & 0   \\ 
 & Boredom 					& 0  & 0  & 15 & \textbf{68}  & 17  & 0  & 0  & 14 & \textbf{71}  & 15  & 0  & 5  & 31 & \textbf{33}  & 31  & 0  & 8  & 34 & \textbf{28}  & 30  & 3  & 6  & 26 & \textbf{51}  & 14 	\\ 
 & Sadness 					& 0  & 0 & 0 & 19 & \textbf{81}  	& 0  & 0 & 0 & 18 & \textbf{82}  	& 0  & 3 & 8 & 15 & \textbf{74}  	& 0  & 5 & 11 & 25 & \textbf{59}  	& 4  & 0 & 6 & 15 & \textbf{75}  	\\ \cline{3-27}
&  & \multicolumn{5}{c||}{\rule{0pt}{7pt} Average: \textbf{74.6}}   & \multicolumn{5}{c||}{Average: \textbf{76.8}}  & \multicolumn{5}{c||}{Average: \textbf{62.6}}  & \multicolumn{5}{c||}{Average: \textbf{55.8}}  & \multicolumn{5}{c|}{Average: \textbf{47.4}}\\ \hline
\hline
\multirow{8}{*}{\rotatebox[origin=c]{90}{SVM Classifier}} & \rule{0pt}{7pt}Actual & \multicolumn{5}{c||}{Classified Emotion}  
& \multicolumn{5}{c||}{Classified Emotion}  & \multicolumn{5}{c||}{Classified Emotion} & \multicolumn{5}{c||}{Classified Emotion} & \multicolumn{5}{c|}{Classified Emotion} \\  
 & Emotion & Ang. & Hap. & Neu. & Bor. & Sad.  & Ang. & Hap. & Neu. & Bor. & Sad.  & Ang. & Hap. & Neu. & Bor. & Sad. & Ang. & Hap. & Neu. & Bor. & Sad.  & Ang. & Hap. & Neu. & Bor. & Sad.  \\ \cline{2-27}  
 & \rule{0pt}{7pt} Anger 	& \textbf{72}  &  28 & 0 & 0 & 0 	& \textbf{73}  &  27 & 0 & 0 & 0 	& \textbf{69}  &  30 & 1 & 0 & 0 	& \textbf{63}  &  30 & 7 & 0 & 0 	& \textbf{20}  &  56 & 24 & 0 & 0	\\ 
 & Happiness 				& 37 &	\textbf{63}  & 0 & 0 & 0 	& 36 &	\textbf{64}  & 0 & 0 & 0 	& 35 &	\textbf{57}  & 8 & 0 & 0 	& 27 &	\textbf{65}  & 8 & 0 & 0 	& 30 &	\textbf{55}  & 10 & 5 & 0 	\\ 
 & Neutral 					& 0 & 0 & \textbf{64}  & 34 & 2 	& 0 & 0 & \textbf{67}  & 23 & 0 	& 0 & 8 & \textbf{56}  & 24 & 12 	& 0 & 20 & \textbf{43}  & 25 & 12 	& 13 & 36 & \textbf{20}  & 31 & 0 	\\ 
 & Boredom 					& 0  & 0  & 20 & \textbf{51} & 29 	& 0  & 0  & 19 & \textbf{52}  & 29 	& 0  & 9  & 28 & \textbf{27}  & 36 	& 0  & 11  & 37 & \textbf{19}  & 33 & 4  & 7  & 27 & \textbf{47}  & 15 	\\ 
 & Sadness 					& 0  & 0 & 0 & 29 & \textbf{71} 	& 0  & 0 & 0 & 27 & \textbf{73}		& 0  & 2 & 10 & 20 & \textbf{68} 	& 0  & 12 & 24 & 35 & \textbf{29} 	& 7	  & 7 & 0 & 17 & \textbf{69} 	\\ \cline{3-27} 
&  & \multicolumn{5}{c||}{\rule{0pt}{7pt} Average: \textbf{64.2}}   & \multicolumn{5}{c||}{Average: \textbf{65.8}}  & \multicolumn{5}{c||}{Average: \textbf{55.4}}  & \multicolumn{5}{c||}{Average: \textbf{43.8}}  & \multicolumn{5}{c|}{Average: \textbf{42.2}}\\ \hline
\end{tabular}
\end{center}
\label{tab:hhhc_eemd_5emos}
\vspace{-.4cm}
\end{table*}

Extensive experiments are carried out to evaluate the proposed HHHC acoustic feature. Figure \ref{fig:modelo} illustrates the classification system used in the experiments. 
Affective models are generated in the training phase after pre-processing and feature extraction. During tests, for each voiced speech signal, the extracted feature vector is compared to each model. 
The leave-one-speaker-out (LOSO) \mbox{methodology}~\cite{schuller2009acoustic} is adopted to achieve speaker independence. For all databases, the modelling of each affective state is conducted with 32 s randomly selected from the training data. Test experiments are applied to 800 ms speech segments of each emotion of the testing speaker. The detection of emotional content in instances which last less than 1~s is suitable for real-life situations~\cite{tahon2016towards}.

The $\alpha$-GMM is performed with five values of $\alpha$: $-1$ (classical GMM), $-2$, $-4$, $-6$ and $-8$. Affective models are composed of 32 Gaussian \mbox{densities} with diagonal covariance matrices. The HMM is implemented using the HTK toolkit~\cite{young2002htk} with the left-to-right topology. For each affective condition, it is  used five HMM states with one single Gaussian mixture per state. The SVM implementation is carried out with the LIBSVM~\cite{chang2011libsvm}, using the ``one-versus-one'' strategy. The search for the optimal hyperplane is conducted in a grid-search procedure for the RBF kernel, with the controlling parameters being evaluated for c $\in$ (0, 10) and $\gamma \in$ (0, 1). The DNNs consider multilayer perceptrons with three hidden layers~\cite{wang2014training}. The networks are trained with the standard backpropagation algorithm with dropout regularization (dropout rate 0.2). It is not used any unsupervised pretraining. The momentum rate used is 0.5. Sigmoid activation functions are used in the output layer, while linear functions are used for the rest.
CNNs and CRNNs are implemented with three convolutional layers followed by max pooling operation 
with (2,2,2) and (5,4,2) pool arrangements, respectively \cite{cakir2017}. 
A single recurrent layer is used to compose the CRNN.

In order to verify the improvement in classification rates for emotion recognition, the proposed HHHC vector is experimented as complementary to collections of features such as GeMAPS~\cite{eyben2016geneva}. For this purpose, binary arousal and valence classification is carried out by using the SVM classifier.

\vspace{-.2cm}
\subsection{Speech Emotion Databases}
\vspace{-.1cm}

Three databases are considered in the experiments: EMO-DB \cite{berlinDatabase}, IEMOCAP (Interactive Emotional Dyadic Motion Capture)~\cite{busso2008iemocap}, and SEMAINE (Sustained Emotionally \mbox{colored} Machine-human Interaction using Nonverbal Expression)~\cite{mckeown2012semaine}.
Only the voiced segments of speech are considered in the experiments. For this purpose, the pre-processing step selects frames of 16 ms with high energy and low zero crossing rate. The sampling rate used  for all databases is 8~kHz. 

EMO-DB consists of ten actors~(5 women and 5 men) that uttered ten sentences in German with archetypical emotions. In this work, five emotional states are considered: Anger, Happiness, Neutral, Boredom and Sadness. Although EMO-DB comprises seven emotions~(including disgust and fear), the experiments with five of them are carried out in order to show the power of an acoustic feature in characterize emotions that are naturally recognized by humans. Thus, five emotions were chosen to show the effectiveness of the HHHC vector. The entire set of voiced speech samples for each emotional state has 40~s.

IEMOCAP is composed of conversations of both scripted and spontaneous scenarios in English language. Ten actors~(5 women and 5 men) were recorded in dyadic sessions in order to facilitate a more natural interaction of the targeted emotion. Since it is analyzed short emotional instances in the test phase, it is used a portion of the IEMOCAP database, although it comprises 12 hours of recordings. It is considered four emotional states: Anger, Happiness, Neutral and Sadness.  A total of 10 minutes of voiced content from each emotional state is used in the experiments, where it is considered 5 minutes of both tasks~(scripted and spontaneous scenarios).

The SEMAINE database features 150 participants~(undergraduate and postgraduate students from eight different countries). The Sensitive Artificial Listener~(SAL) scenario was used in conversations in English. Interactions involve a ``user'' (human) and an \mbox{``operator''} (either a machine or a person simulating a machine). In this work, it is \mbox{considered} recordings from ten participants~(5 women and 5 men). From 27 categories~(styles), 4 emotional states were selected: Anger, Happiness, Amusement and Sadness.  The set of voiced speech samples for each emotional state has 90~s.

\vspace{-.2cm}
\subsection{Extracted Features}

6-dimensional HHHC vectors are extracted according to the procedure presented in the Section~II-A. In the EEMD-based analysis, it is experimented 11 Gaussian noise levels, considering the noise standard deviation (std) in the range $\left[0.005, 0.1 \right]$. The robustness of the HHHC is also verified using the INS in the feature vector~(HHHC+INS). For each IMF, the INS values are computed with ten different observation scales, $T_{h}/T \in \left[0.0015, 0.5\right]$. 

For the performance comparison and feature fusion, MFCC, TEO-CB-Auto-Env and pH vector are used in the experiments.
12-dimensional MFCC vectors are obtained from speech frames of 25 ms, with a frame rate of 10 ms. For the TEO-CB-Auto-Env~(TEO feature), vectors with 16 coefficients are extracted from 75~ms speech samples, with 50\% overlapping. The estimation of the pH feature is conducted in frames of 50 ms, every 10 ms, using the Daubechies wavelet filters with 12 coefficients~(2-12 scales). Fusion procedures are carried out for an improvement provided by the proposed HHHC in the recognition rates of the baseline features.

\begin{table*}[!t]
\renewcommand{\arraystretch}{1.05}
\centering 
\fontsize{7}{7}\selectfont
\vspace{-.4cm}
\setlength{\tabcolsep}{4pt}
\caption{Accuracy Rates (\%) of 4 Emotional States with the HHHC and baseline features for IEMOCAP.}
\vspace{-.4cm}
\begin{center}
\begin{tabular}{|c|c|c|c|c|c||c|c|c|c||c|c|c|c||c|c|c|c||c|c|c|c|}
\cline{3-22} 
\multicolumn{2}{c|}{\rule{0pt}{7pt}} & \multicolumn{4}{c||}{HHHC feature}  & \multicolumn{4}{c||}{HHHC + INS} & \multicolumn{4}{c||}{pH feature} 
& \multicolumn{4}{c||}{MFCC feature} & \multicolumn{4}{c|}{TEO feature} \\ 
\hline
\multirow{7}{*}{\rotatebox[origin=c]{90}{$\alpha$-GMM Classifier}} & \rule{0pt}{7pt}Actual & \multicolumn{4}{c||}{Classified Emotion}  & \multicolumn{4}{c||}{Classified Emotion}  
& \multicolumn{4}{c||}{Classified Emotion}  & \multicolumn{4}{c||}{Classified Emotion}  & \multicolumn{4}{c|}{Classified Emotion} \\ 
 & Emotion & Ang. & Hap. & Neu.  & Sad.  & Ang. & Hap. & Neu.  & Sad.  & Ang. & Hap. & Neu. & Sad.  & Ang. & Hap. & Neu. & Sad.  & Ang. & Hap. & Neu. & Sad.  \\ \cline{2-22}   
 & \rule{0pt}{7pt} Anger 	& \textbf{66}  & 23 & 9 & 2 	& \textbf{68}  &	23 & 9 & 0 	& \textbf{59}  &	24 & 13 & 4 & \textbf{59}  & 16 & 15 & 10 	& \textbf{40} & 25 & 24 & 11 	\\ 
 & Happiness 				& 26  &	\textbf{55} & 15 & 4 	& 26  &	\textbf{57} & 15 & 2 	& 28  &	\textbf{47} & 17 & 8 	& 28  &	\textbf{43} & 20 & 9 	& 33  &	\textbf{36} & 21 & 10 	\\ 
 & Neutral 					& 10  &	 12 & \textbf{61} & 17 	& 9  &	 11 & \textbf{63} & 17 	& 12  &	 15 & \textbf{52} & 21 	& 16  &	 11 & \textbf{47} & 26 	& 7  &	 24 & \textbf{37} & 32 	\\ 
 & Sadness 					& 7  &	9 & 22 & \textbf{62} 	& 6  &	9 & 22 & \textbf{63} 	& 9  &	13 & 25 & \textbf{53} 	& 9  &	11 & 26 & \textbf{54} 	& 8  &	5 & 23 & \textbf{64} 	\\ \cline{3-22} 
&  & \multicolumn{4}{c||}{\rule{0pt}{7pt} Average: \textbf{61.0}}   & \multicolumn{4}{c||}{Average: \textbf{62.8}}  & \multicolumn{4}{c||}{Average: \textbf{52.8}}  & \multicolumn{4}{c||}{Average: \textbf{50.8}}  & \multicolumn{4}{c|}{Average: \textbf{44.2}} \\ \hline
\hline
\multirow{7}{*}{\rotatebox[origin=c]{90}{HMM Classifier }} & \rule{0pt}{7pt}Actual & \multicolumn{4}{c||}{Classified Emotion}  & \multicolumn{4}{c||}{Classified Emotion}  
& \multicolumn{4}{c||}{Classified Emotion}  & \multicolumn{4}{c||}{Classified Emotion}  & \multicolumn{4}{c|}{Classified Emotion} \\ 
 & Emotion & Ang. & Hap. & Neu.  & Sad.  & Ang. & Hap. & Neu.  & Sad.  & Ang. & Hap. & Neu. & Sad.  & Ang. & Hap. & Neu. & Sad.  & Ang. & Hap. & Neu. & Sad.  \\ \cline{2-22}   
 & \rule{0pt}{7pt} Anger 	& \textbf{55}  & 28 & 12 & 5 	& \textbf{58}  & 28 & 13 & 1 	& \textbf{57}  & 26 & 13 & 4 	& \textbf{50}  & 19 & 18 & 13 	& \textbf{37}  & 26 & 25 & 12 	\\              
 & Happiness 				& 31  &	\textbf{45} & 19 & 5 	& 30  &	\textbf{48} & 18 & 4 	& 33  &	\textbf{42} & 17 & 8 	& 30  &	\textbf{37} & 22 & 11 	& 35  &	\textbf{31} & 22 & 12 	\\              
 & Neutral 					& 10  &	15 & \textbf{54} & 21 	& 11  &	13 & \textbf{57} & 19 	& 12  &	15 & \textbf{49} & 24 	& 16  &	12 & \textbf{44} & 28 	& 8  &	25 & \textbf{33} & 34 	\\              
 & Sadness 					& 7  &	12 & 27 & \textbf{54} 	& 6  &	10 & 26 & \textbf{58} 	& 10  &	14 & 27 & \textbf{49} 	& 10  &	12 & 28 & \textbf{50} 	& 9  &	8 & 24 & \textbf{59} 	\\ \cline{3-22} 
&  & \multicolumn{4}{c||}{\rule{0pt}{7pt} Average: \textbf{52.0}}   & \multicolumn{4}{c||}{Average: \textbf{55.3}}  & \multicolumn{4}{c||}{Average: \textbf{49.3}}  & \multicolumn{4}{c||}{Average: \textbf{45.3}}  & \multicolumn{4}{c|}{Average: \textbf{40.0}} \\ \hline
\hline
\multirow{7}{*}{\rotatebox[origin=c]{90}{SVM Classifier }} & \rule{0pt}{7pt}Actual & \multicolumn{4}{c||}{Classified Emotion}  & \multicolumn{4}{c||}{Classified Emotion}  
& \multicolumn{4}{c||}{Classified Emotion}  & \multicolumn{4}{c||}{Classified Emotion}  & \multicolumn{4}{c|}{Classified Emotion} \\ 
 & Emotion & Ang. & Hap. & Neu.  & Sad.  & Ang. & Hap. & Neu.  & Sad.  & Ang. & Hap. & Neu. & Sad.  & Ang. & Hap. & Neu. & Sad.  & Ang. & Hap. & Neu. & Sad.  \\ \cline{2-22}   
 & \rule{0pt}{7pt} Anger 	& \textbf{49}  &	31 & 14 & 6 & \textbf{51}  &	31 & 14 & 4 & \textbf{49}  &30 & 15 & 6 	& \textbf{40}  & 22 & 23 & 15 	& \textbf{27}  & 30 & 29 & 14 	\\              
 & Happiness 				& 30  &	\textbf{35} & 28 & 7 	& 30  &	\textbf{38} & 27 & 5 	& 29  &	\textbf{30} & 26 & 15 	& 32  &	\textbf{32} & 24 & 12 	& 37  &	\textbf{25} & 24 & 14 	\\              
 & Neutral 					& 15  &	20 & \textbf{39} & 26 	& 15  &	19 & \textbf{40} & 26 	& 17  &	24 & \textbf{32} & 27 	& 18  &	15 & \textbf{31} & 36 	& 9  &	27 & \textbf{26} & 38 	\\              
 & Sadness 					& 7  &	14 & 33 & \textbf{46} 	& 7  &	14 & 32 & \textbf{47} 	& 12  &	15 & 33 & \textbf{40} 	& 13  &	15 & 31 & \textbf{41} 	& 9  &	9 & 27 & \textbf{55} 	\\ \cline{3-22} 
&  & \multicolumn{4}{c||}{\rule{0pt}{7pt} Average: \textbf{42.3}}   & \multicolumn{4}{c||}{Average: \textbf{44.0}}  & \multicolumn{4}{c||}{Average: \textbf{37.8}}  & \multicolumn{4}{c||}{Average: \textbf{36.0}}  & \multicolumn{4}{c|}{Average: \textbf{33.3}} \\ \hline
\end{tabular}
\end{center}
\vspace{-.4cm}
\label{tab:hhhc_eemd_4emosIEMOCAP}
\end{table*}

\section{Results}
\label{sec:result}

This Section presents accuracies results obtained in speech emotion classification.
For this purpose, confusion matrices are obtained considering the proposed HHHC and baseline features.
Tables~\ref{tab:hhhc_eemd_5emos},~\ref{tab:hhhc_eemd_4emosIEMOCAP} and~\ref{tab:hhhc_eemd_4emosSEMAINE} present \mbox{accuracies} achieved for the EMO-DB, IEMOCAP and SEMAINE databases, respectively. They show confusion matrices obtained with $\alpha$-GMM, HMM and SVM classifiers for the HHHC, HHHC+INS, and baseline features. The EMD-based HHHC already outperforms competing attributes. However, the EEMD-based approach reaches superior accuracies.  Results for HHHC are achieved with the EEMD-based approach considering low Gaussian noise \mbox{level~(0.005$\leq$ std $\leq$0.02)}.

\begin{figure}[!t]
\centering
\includegraphics[width=7.cm]{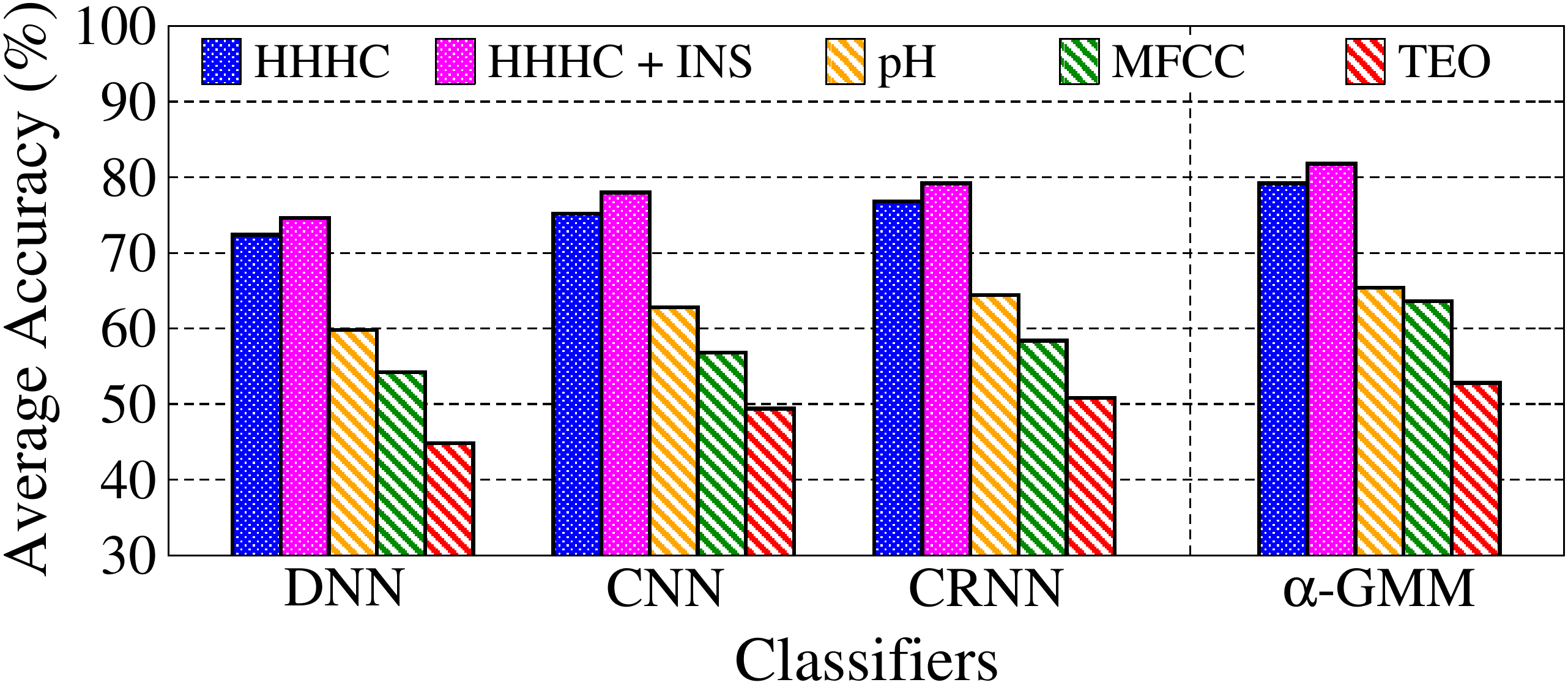}
\vspace{-.1cm}
\caption{Average accuracies of EMO-DB obtained with $\alpha$-GMM and Neural Network classifiers.}
\vspace{-.2cm}
\label{fig:nn_emodb}
\end{figure}

\begin{figure}[!t]
\centering
\includegraphics[width=7.cm]{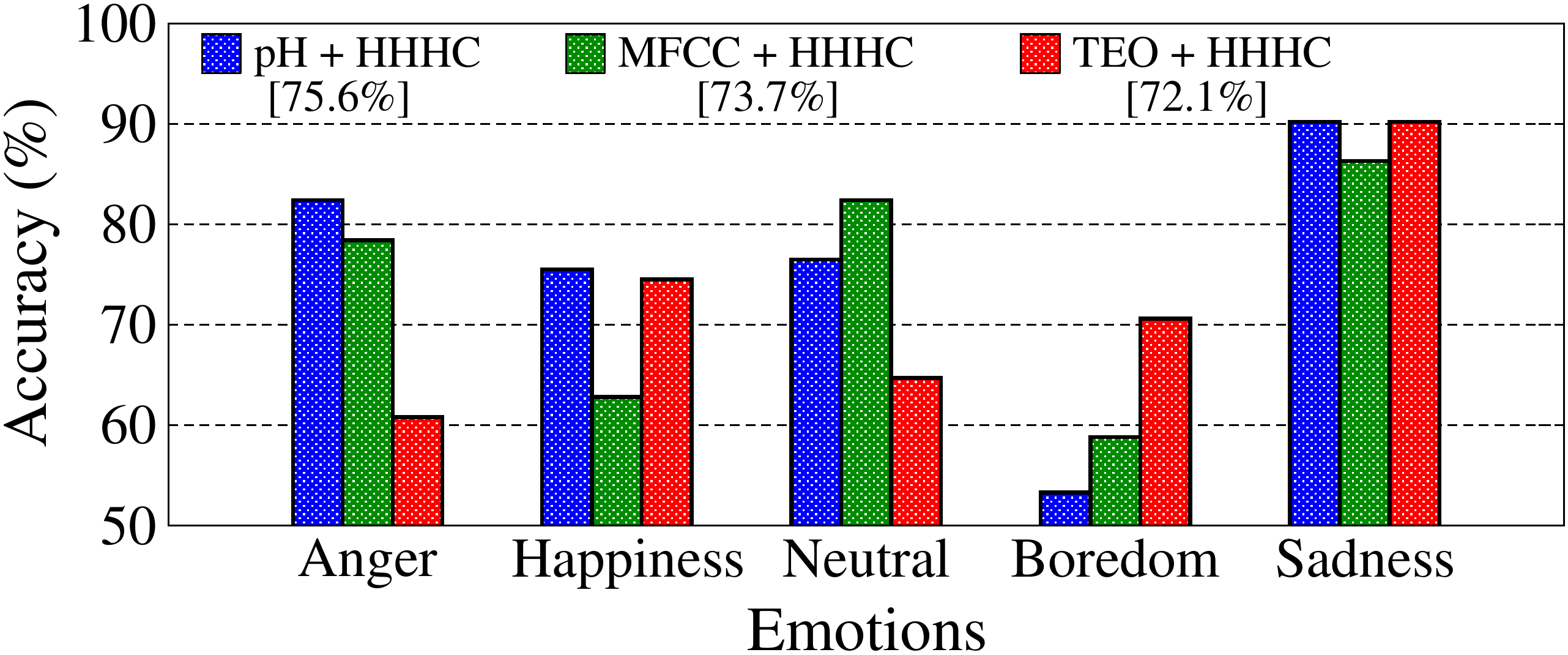}
\vspace{-.1cm}
\caption{Classification accuracies with feature fusion and $\alpha$-GMM classifier of emotional states from EMO-DB.}
\vspace{-.2cm}
\label{fig:feat_fusion_emodb}
\end{figure}

\subsection{Results with EMO-DB}

For the $\alpha$-GMM, the proposed HHHC feature achieves the best average accuracy~(79.2\%) with three values of $\alpha$~($-4$, $-6$ and $-8$). This value is greater than the average accuracy achieved with pH for $\alpha=-2$~(65.4\%). The HHHC also outperforms in 15.6~p.p. the average accuracy of MFCC~(63.6\%), and reaches 26.4~p.p. over the TEO feature~(52.8\%). The INS information contributes for an increasing of more than 2~p.p. over the HHHC. The HHHC enables almost 60.0\% of recognition for each considered emotional state using $\alpha$-GMM. For all considered feature sets, the $\alpha$-GMM (including the original GMM) outperforms the HMM and SVM classifiers. 

Figure \ref{fig:nn_emodb} presents the average classification accuracies obtained with the proposed and baseline features 
considering the Neural Network classifiers.
Average results obtained with the $\alpha$-GMM are also shown in Figure \ref{fig:nn_emodb}.
Note that HHHC and HHHC+INS achieve the best results for all classifiers.
For the CRNN, which outperforms DNN and CNN, HHHC leads to an improvement of 12.4 p.p. over pH: from 64.4\% to 76.8\%.
For this classifier, the average accuracy obtained with HHHC+INS achieves 79.2\%, i.e., 2.4 p.p. higher than HHHC.
It can also be noticed that the introduced $\alpha$-GMM achieves the best classification accuracies for all features sets.
For HHHC+INS features, for example, the average accuracy with $\alpha$-GMM is 2.6 p.p. greater than CRNN.

Figure \ref{fig:feat_fusion_emodb} shows the identification accuracy with $\alpha$-GMM for the feature fusion between HHHC and competing features. 
The best average accuracy attained with the pH+HHHC fusion (75.6\% with $\alpha=-6$) is 10.2 p.p. higher than that achieved with pH only~(65.4\%). The MFCC+HHHC fusion reaches the best accuracy~(73.7\%) with $\alpha=-8$. It means that HHHC increases  in almost 10 p.p. the recognition rate provided by the MFCC feature. About the TEO+HHHC fusion, the best average accuracy is 72.1\% with $\alpha=-6$ and $\alpha=-8$. This means an improvement of 19.2 p.p. provided by the HHHC for the TEO-based feature.

\begin{figure}[!t]
\centering
\includegraphics[width=7.cm]{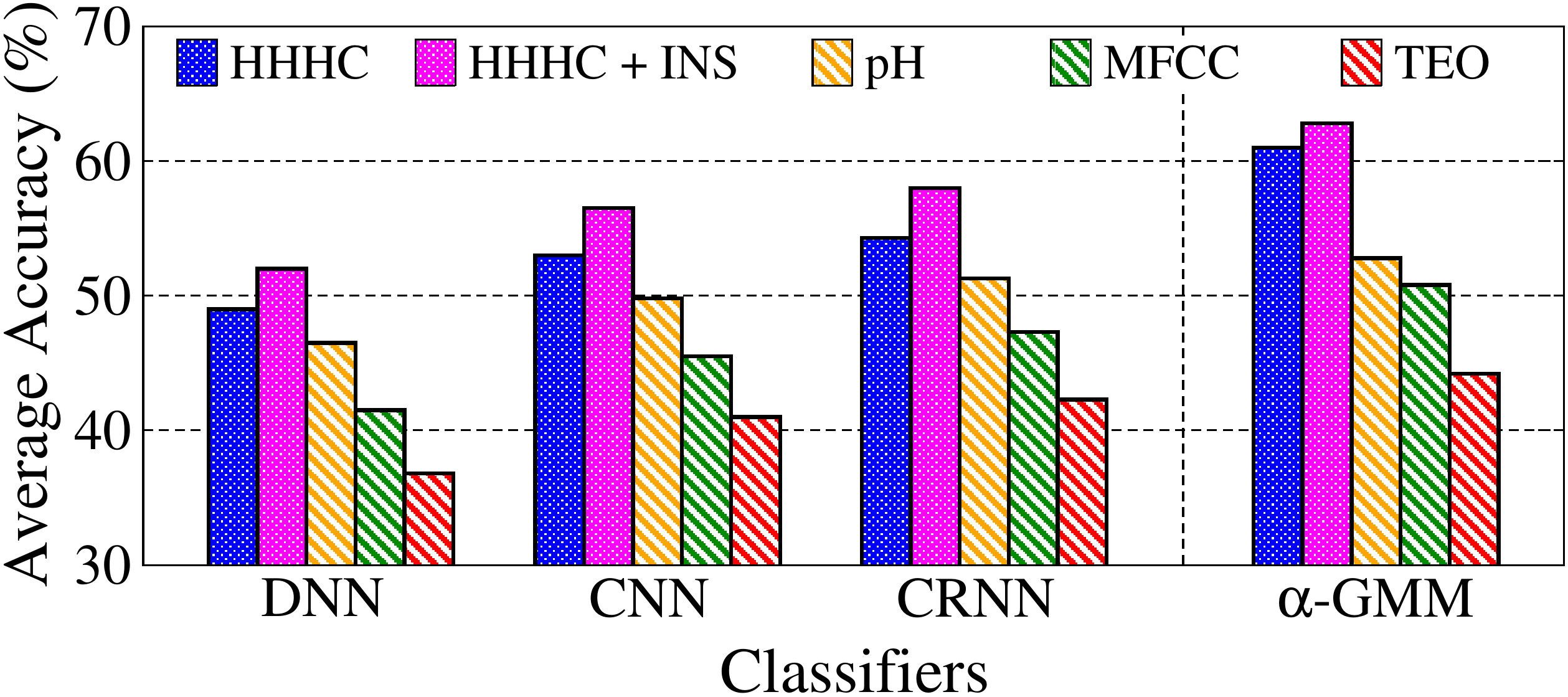}
\vspace{-.1cm}
\caption{Average accuracies of IEMOCAP obtained with $\alpha$-GMM and Neural Network classifiers.}
\vspace{-.2cm}
\label{fig:nn_iemocap}
\end{figure}

\begin{figure}[!t]
\centering
\includegraphics[width=7.cm]{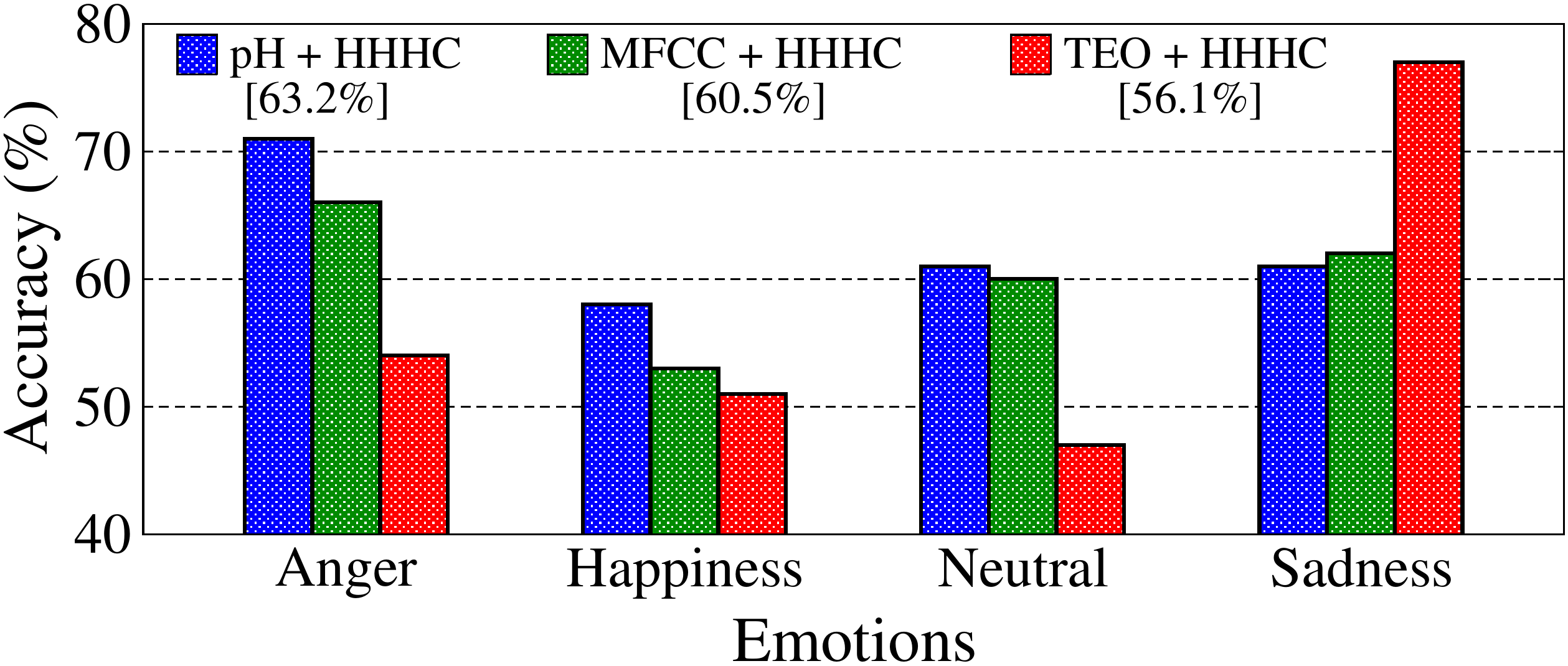}
\vspace{-.1cm}
\caption{Classification accuracies with feature fusion and $\alpha$-GMM classifier of emotional states from IEMOCAP.}
\vspace{-.2cm}
\label{fig:feat_fusion_iemocap}
\end{figure}

\vspace{-.2cm}
\subsection{Results with IEMOCAP}

\begin{table*}[!t]
\renewcommand{\arraystretch}{1.05}
\centering 
\fontsize{7}{7}\selectfont
\vspace{-.4cm}
\setlength{\tabcolsep}{4pt}
\caption{Accuracy Rates (\%) of 4 Emotional States with the HHHC and baseline features for SEMAINE.}
\vspace{-.4cm}
\begin{center}
\begin{tabular}{|c|c|c|c|c|c||c|c|c|c||c|c|c|c||c|c|c|c||c|c|c|c|}
\cline{3-22} 
\multicolumn{2}{c|}{\rule{0pt}{7pt}} & \multicolumn{4}{c||}{HHHC feature}  & \multicolumn{4}{c||}{HHHC + INS} & \multicolumn{4}{c||}{pH feature} 
& \multicolumn{4}{c||}{MFCC feature} & \multicolumn{4}{c|}{TEO feature} \\ 
\hline
\multirow{7}{*}{\rotatebox[origin=c]{90}{$\alpha$-GMM Classifier}} & \rule{0pt}{7pt}Actual & \multicolumn{4}{c||}{Classified Emotion}  & \multicolumn{4}{c||}{Classified Emotion}  
& \multicolumn{4}{c||}{Classified Emotion}  & \multicolumn{4}{c||}{Classified Emotion}  & \multicolumn{4}{c|}{Classified Emotion} \\ 
 & Emotion & Ang. & Hap. & Amu.  & Sad.  & Ang. & Hap. & Amu.  & Sad.  & Ang. & Hap. & Amu. & Sad.  & Ang. & Hap. & Amu. & Sad.  & Ang. & Hap. & Amu. & Sad.  \\ \cline{2-22}   
 & \rule{0pt}{7pt} Anger 	& \textbf{50}  & 23 & 20 & 7 	& \textbf{51}  & 23 & 20 & 6 	& \textbf{50}  & 22 & 20 & 8 	& \textbf{42}  & 29 & 16 & 13 	& \textbf{34}  & 24 & 22 & 20 	\\ 
 & Happiness 				& 14  &	\textbf{57} & 25 & 4 	& 14  &	\textbf{59} & 25 & 2 	& 17  & \textbf{51} & 27 & 5 	& 18  &	\textbf{52} & 26 & 4 	& 29  &	\textbf{33} & 29 & 9 	\\ 
 & Amusement 				& 14  &	 26 & \textbf{51} & 9 	& 13  &	 24 & \textbf{55} & 8 	& 16  &	 26 & \textbf{48} & 10 	& 15  &	 30 & \textbf{47} & 8 	& 19  &	 25 & \textbf{35} & 21 	\\ 
 & Sadness 					& 6  &	15 & 19 & \textbf{60} 	& 5  &	15 & 17 & \textbf{63} 	& 8  &	15 & 23 & \textbf{54} 	& 9  &	11 & 25 & \textbf{55} 	& 3  &	16 & 20 & \textbf{61} 	\\ \cline{3-22} 
&  & \multicolumn{4}{c||}{\rule{0pt}{7pt} Average: \textbf{54.5}}   & \multicolumn{4}{c||}{Average: \textbf{57.0}}  & \multicolumn{4}{c||}{Average: \textbf{50.8}}  & \multicolumn{4}{c||}{Average: \textbf{49.0}}  & \multicolumn{4}{c|}{Average: \textbf{40.8}} \\ \hline
\hline
\multirow{7}{*}{\rotatebox[origin=c]{90}{HMM Classifier }} & \rule{0pt}{7pt}Actual & \multicolumn{4}{c||}{Classified Emotion}  & \multicolumn{4}{c||}{Classified Emotion}  
& \multicolumn{4}{c||}{Classified Emotion}  & \multicolumn{4}{c||}{Classified Emotion}  & \multicolumn{4}{c|}{Classified Emotion} \\ 
 & Emotion & Ang. & Hap. & Amu.  & Sad.  & Ang. & Hap. & Amu.  & Sad.  & Ang. & Hap. & Amu. & Sad.  & Ang. & Hap. & Amu. & Sad.  & Ang. & Hap. & Amu. & Sad.  \\ \cline{2-22}   
 & \rule{0pt}{7pt} Anger 	& \textbf{45}  & 26 & 22 & 7 	& \textbf{46}  & 25 & 22 & 7 	& \textbf{45}  & 25 & 22 & 8 	& \textbf{38}  & 31 & 17 & 14 	& \textbf{28}  & 26 & 24 & 22 	\\              
 & Happiness 				& 17  &	\textbf{50} & 28 & 5 	& 17  &	\textbf{53} & 28 & 2 	& 19  &	\textbf{47} & 29 & 5 	& 19  &	\textbf{49} & 28 & 4 	& 30  &	\textbf{31} & 30 & 9 	\\              
 & Amusement				& 14  &	29 & \textbf{48} & 9 	& 13  &	27 & \textbf{51} & 9 	& 16  &	28 & \textbf{45} & 11 	& 16  &	31 & \textbf{42} & 11 	& 20  &	27 & \textbf{31} & 22 	\\              
 & Sadness 					& 8  &	18 & 22 & \textbf{52} 	& 5  &	18 & 22 & \textbf{55} 	& 8  &	18 & 27 & \textbf{47} 	& 10  &	13 & 30 & \textbf{47} 	& 3  &	18 & 24 & \textbf{55} 	\\ \cline{3-22} 
&  & \multicolumn{4}{c||}{\rule{0pt}{7pt} Average: \textbf{48.8}}   & \multicolumn{4}{c||}{Average: \textbf{51.3}}  & \multicolumn{4}{c||}{Average: \textbf{46.0}}  & \multicolumn{4}{c||}{Average: \textbf{44.0}}  & \multicolumn{4}{c|}{Average: \textbf{36.2}} \\ \hline
\hline
\multirow{7}{*}{\rotatebox[origin=c]{90}{SVM Classifier }} & \rule{0pt}{7pt}Actual & \multicolumn{4}{c||}{Classified Emotion}  & \multicolumn{4}{c||}{Classified Emotion}  
& \multicolumn{4}{c||}{Classified Emotion}  & \multicolumn{4}{c||}{Classified Emotion}  & \multicolumn{4}{c|}{Classified Emotion} \\ 
 & Emotion & Ang. & Hap. & Amu.  & Sad.  & Ang. & Hap. & Amu.  & Sad.  & Ang. & Hap. & Amu. & Sad.  & Ang. & Hap. & Amu. & Sad.  & Ang. & Hap. & Amu. & Sad.  \\ \cline{2-22}   
 & \rule{0pt}{7pt} Anger 	& \textbf{39}  & 28 & 24 & 9 	& \textbf{41}  &	28 & 24 & 7 & \textbf{38}  &	29 & 25 & 8 & \textbf{30}  & 34 & 20 & 16 	& \textbf{18}  & 30 & 28 & 24 	\\              
 & Happiness 				& 20  &	\textbf{43} & 32 & 5 	& 19  &	\textbf{45} & 31 & 5 	& 22  &	\textbf{40} & 33 & 5 	& 21  &	\textbf{41} & 33 & 5 	& 33  &	\textbf{22} & 35 & 10 	\\              
 & Amusement				& 16  &	32 & \textbf{43} & 9 	& 15  &	30 & \textbf{44} & 11 	& 18  &	30 & \textbf{39} & 13 	& 18  &	34 & \textbf{35} & 13 	& 21  &	29 & \textbf{24} & 26 	\\              
 & Sadness 					& 9  & 20 & 25 & \textbf{46} 	& 7  & 20 & 26 & \textbf{47} 	& 9  &	20 & 31 & \textbf{40}  	& 11  &	15 & 35 & \textbf{39} 	& 3  &	21 & 29 & \textbf{47} 	\\ \cline{3-22} 
&  & \multicolumn{4}{c||}{\rule{0pt}{7pt} Average: \textbf{42.8}}   & \multicolumn{4}{c||}{Average: \textbf{44.3}}  & \multicolumn{4}{c||}{Average: \textbf{39.3}}  & \multicolumn{4}{c||}{Average: \textbf{36.3}}  & \multicolumn{4}{c|}{Average: \textbf{27.8}} \\ \hline
\end{tabular}
\end{center}
\vspace{-.5cm}
\label{tab:hhhc_eemd_4emosSEMAINE}
\end{table*}

It can be seen from Table~\ref{tab:hhhc_eemd_4emosIEMOCAP} that, for all considered feature sets, 
the $\alpha$-GMM achieves superior accuracies over the HMM and the SVM classifiers. 
Only HHHC and HHHC+INS reach average \mbox{accuracies} over 60.0\%. 
These values are achieved using the $\alpha$-GMM with $\alpha=-8$. In comparison to baseline features, HHHC obtained an average accuracy  8~p.p.  over the pH vector~($\alpha=-8$), 10~p.p.  over the \mbox{MFCC~($\alpha=-4$)} and 15~p.p over  the TEO-based feature~($\alpha=-6$). For each considered emotional state, the $\alpha$-GMM approach achieves more than 50.0\% accuracies with HHHC. Furthermore, $\alpha$-GMM provides an improved performance with baseline features, in comparison to HMM and SVM approaches.

Figure \ref{fig:nn_iemocap} presents the average classification accuracies of IEMOCAP
considering $\alpha$-GMM and Neural Network classifiers.
As in the EMO-DB, HHHC outperforms the pH, MFCC and TEO features for all classifiers.
For the CRNN, HHHC achieves an average accuracy of 54.3\%, which is 3.0 p.p., 7.0 p.p., and 12.0 p.p. greater than
pH, MFCC, and TEO, respectively.
Moreover, HHHC+INS leads to the best results for all scenarios. 
The $\alpha$-GMM also outperforms the competing classifiers for all features sets.

\begin{figure}[!t]
\centering
\includegraphics[width=7.cm]{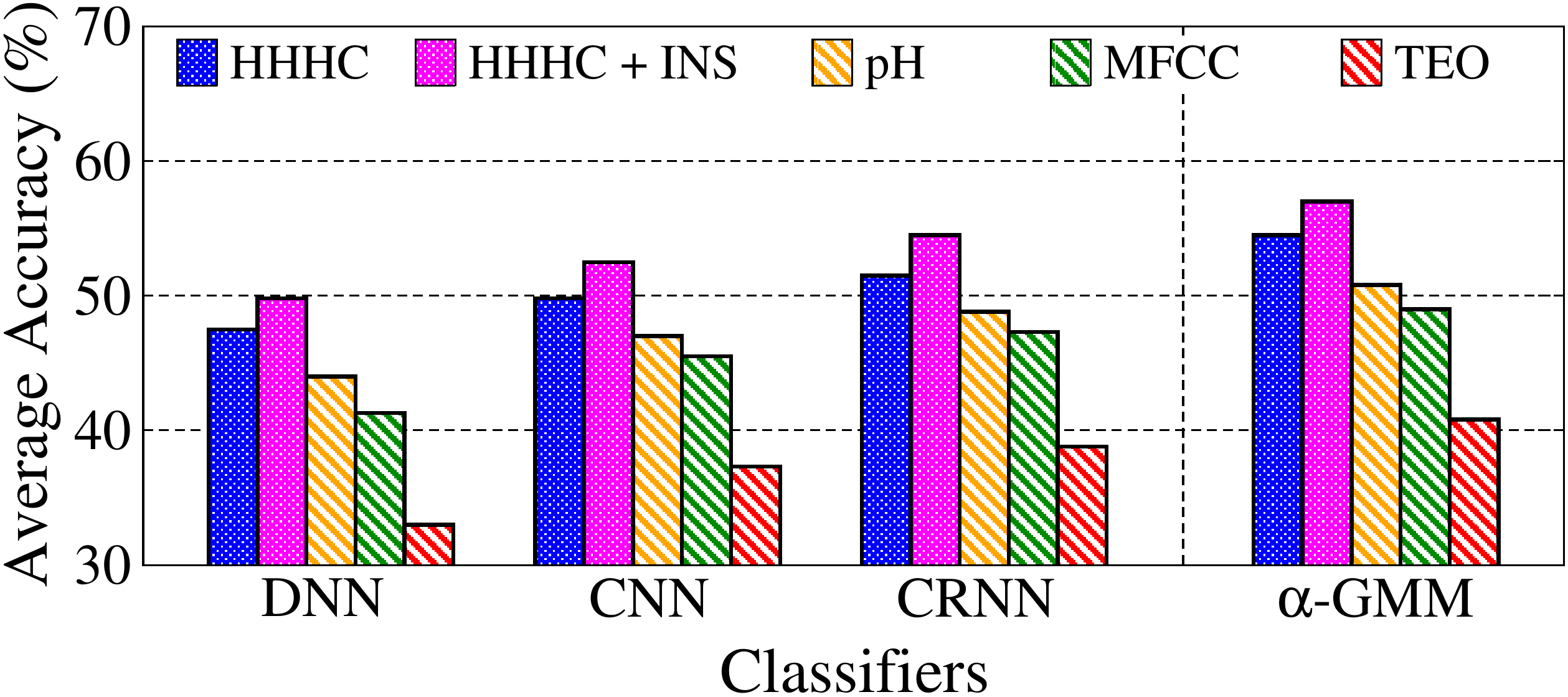}
\vspace{-.3cm}
\caption{Average accuracies of SEMAINE obtained with $\alpha$-GMM and Neural Network classifiers.}
\vspace{-.2cm}
\label{fig:nn_semaine}
\end{figure}

Figure \ref{fig:feat_fusion_iemocap} depicts results of the feature fusion using the $\alpha$-GMM for the HHHC and baseline features in the IEMOCAP database. The pH+HHHC fusion achieves an accuracy of 63.2\%~($\alpha=-8$), which outperforms both pH~(52.8\%) and HHHC+INS~(62.8\%). The fusion of Hurst-based features~(pH+HHHC) indicates that the relation between $H$ and the excitation source enables a high performance in the separation of basic emotions. As for the MFCC+HHHC fusion, HHHC leads to the MFCC an improvement in the average accuracy from 50.8\% to 60.5\%~($\alpha=-4$). Considering the TEO+HHHC fusion, the best average \mbox{accuracy}~(56.1\%) is achieved with $\alpha=-4$, which is 11.9 p.p. higher than that obtained with the TEO-based feature only.

\subsection{Results with SEMAINE}

\begin{figure}[!t]
\centering
\includegraphics[width=7.cm]{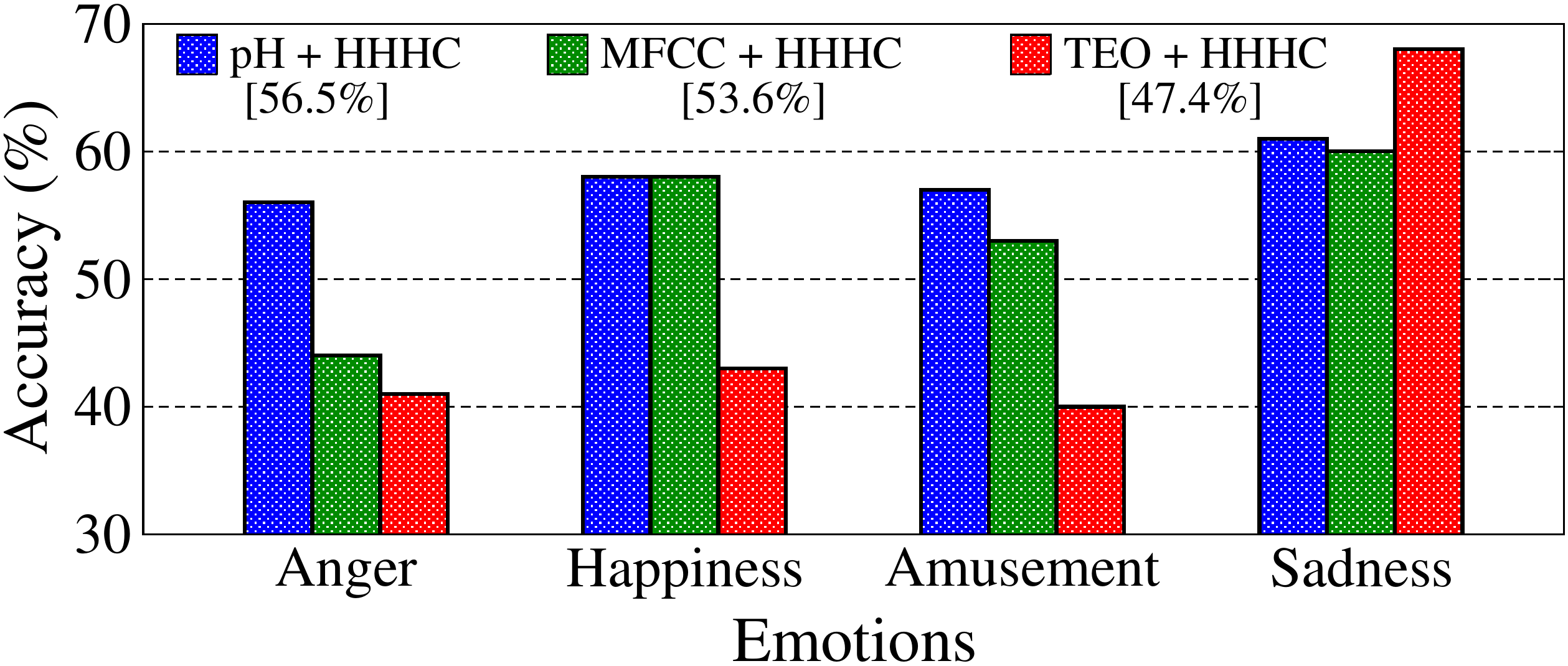}
\vspace{-.3cm}
\caption{Classification accuracies with feature fusion and $\alpha$-GMM classifier of emotional states from SEMAINE.}
\vspace{-.2cm}
\label{fig:feat_fusion_semaine}
\end{figure}

The best average accuracy is achieved with HHHC and HHHC+INS (refer to Table~\ref{tab:hhhc_eemd_4emosSEMAINE}): 54.5\% and 57.0\%, respectively, using $\alpha$-GMM with $\alpha=-6$. These results are greater than 50.8\% for pH~($\alpha=-4$), 49.0\% for the MFCC~($\alpha=-6$), and 40.8\% for TEO-based feature~($\alpha=-8$). An important issue on the SEMAINE database is mainly concerned to the Happiness and Amusement states recognition. Although these emotions present similar behavior, the HHHC shows to be able to recognize both of them with an accuracy over 50.0\% in the classification provided by the $\alpha$-GMM. For baseline features, the $\alpha$-GMM reaches more than 4~p.p. over HMM and 10~p.p. over SVM. The $\alpha$-GMM outperforms HMM and SVM for all considered emotional states. According to the average classification results shown in Figure \ref{fig:nn_semaine}, 
$\alpha$-GMM also outperforms the competing DNN, CNN and CRNN classifiers.
For these classifiers, HHHC and HHHC+INS also achieve the best average results.

The best recognition rates on the feature fusion task with the HHHC and the baseline features using $\alpha$-GMM are shown in Figure \ref{fig:feat_fusion_semaine}. The pH+HHHC fusion attains an average accuracy of 56.5\%, which represents an improvement over pH and HHHC features.  With the MFCC+HHHC feature fusion, it is observed an enhancement from 49.0\% to 53.6\% in the recognition rate, with $\alpha=-6$. The HHHC provides an improvement of more than 6~p.p. when compared to the TEO-based feature~(47.4\%, $\alpha=-8$).
The proposed feature is also very promising for discriminant learning strategies \cite{zhang2018speech} applied to DNN and Deep Convolutional Neural Networks (DCNN) methods for 
speech emotion classification.

\vspace{-.2cm}
\subsection{HHHC Complementarity Aspect}\label{gemapssec}

In order to evaluate the complementarity of the HHHC feature vector to collections of features sets, binary arousal and valence emotion classification are carried out considering all emotions of EMO-DB. The GeMAPS feature set and its extended version (eGeMAPS)~\cite{eyben2016geneva} are adopted for this purpose. The experimental setup is similar to~\cite{eyben2016geneva} with LOSO cross-validation with eight folds, where the speaker IDs are randomly arranged into eight speaker groups. The SVM method is applied for the classification procedure with the LIBSVM toolkit and the same parameters presented in Section~\ref{sec:exp}. Table~\ref{tabgemaps} shows results of UAR (Unweighted \mbox{Average} Recall) obtained from experiments with GeMAPS, eGeMAPS, HHHC, HHHC+INS, and the feature fusion of the proposed acoustic feature with the comparative feature sets. Note that, for arousal evaluation, GeMAPS and eGeMAPS reach more than 93\% UAR while HHHC and HHHC+INS achieve 80.5\% and 83.2\%, respectively. While the standard feature sets needs 62 and 88 features (GeMAPS and eGeMAPS, respectively) for this result, HHHC shows interesting accuracy for a low dimensional feature. However, HHHC and HHHC+INS contribute for an improvement in the UAR obtained with GeMAPS and eGeMAPS. For instance, eGeMAPS+HHHC+INS reaches 98.4\% UAR. In valence classification, HHHC and HHHC+INS also contribute to the feature sets. GeMAPS performance is improved from 74.4\% to 80.4\% with HHHC+INS, while eGeMAPS reaches 82.1\% with this fusion. This experiment demonstrates the complementarity potential of the HHHC to the GeMAPS and eGeMAPS features sets. 

\begin{table}[!t]
\caption{Classification of Binary Arousal and Valence for EMO-DB.}
\vspace{-.4cm}
\footnotesize
\begin{center}
\begin{tabular}{|l||c|c|}
\hline
\multirow{2}{*}{Feature Set} & \multicolumn{2}{c|}{\rule{0pt}{7pt}UAR (\%) with SVM} \\ 
    & Arousal & Valence  \\ \hline   
HHHC & 80.5  &	67.8 \\ 
HHHC+INS & 83.2  &	69.9  \\ 
GeMAPS & 93.2  &	 74.4  \\ 
eGeMAPS & 93.9  &	 74.8  \\ 
GeMAPS+HHHC & 96.1  &	 79.1  \\ 
GeMAPS+HHHC+INS & \textbf{97.6}  &	 \textbf{80.4}  \\ 
eGeMAPS+HHHC & 96.7  &	 81.3  \\ 
eGeMAPS+HHHC+INS & \textbf{98.4}  &	\textbf{82.1}  \\ \hline
\end{tabular}
\end{center}
\label{tabgemaps}
\end{table}

\vspace{-.2cm}
\section{Conclusion}
\label{sec:conclusion}

This work introduced the HHHC nonlinear vocal source feature vector for speech emotion classification. The INS was used as dynamic information for the HHHC vector. Furthermore, the $\alpha$-GMM approach was proposed for this classification task.  It was compared to HMM, SVM, DNN, CNN, and CRNN. The best \mbox{average} classification accuracies were obtained using the $\alpha$-GMM. In comparison to baseline features, HHHC obtained superior accuracy considering three different databases. On the feature fusion, HHHC provides an improved performance for all considered baseline features. As for the EMO-DB, the \mbox{highest} classification accuracy was 81.8\% with HHHC+INS. For the IEMOCAP database, it was reached an average accuracy of 63.2\% with pH+HHHC. In the SEMAINE context, the best average accuracy was 57.0\% with HHHC+INS. 
The superior performance of the proposed feature showed that the HHHC is very promising for affective state representation and for classification tasks. Also, the HHHC complementarity to GeMAPS features set was verified by the improvement in the recognition rates in binary arousal and valence emotion classification.

\bibliographystyle{IEEEtran} 
\bibliography{ref2017} 

\end{document}